\documentclass[]{article}
\renewcommand{\arraystretch}{1.3}

\catcode`\@=11
\def\marginnote#1{}

\newcount\hour
\newcount\minute
\newtoks\amorpm
\hour=\time\divide\hour by60
\minute=\time{\multiply\hour by60 \global\advance\minute by-\hour}
\edef\standardtime{{\ifnum\hour<12 \global\amorpm={am}%
        \else\global\amorpm={pm}\advance\hour by-12 \fi
        \ifnum\hour=0 \hour=12 \fi
        \number\hour:\ifnum\minute<10 0\fi\number\minute\the\amorpm}}
\edef\militarytime{\number\hour:\ifnum\minute<10 0\fi\number\minute}

\def\draftlabel#1{{\@bsphack\if@filesw {\let\thepage\relax
      \xdef\@gtempa{\write\@auxout{\string
          \newlabel{#1}{{\@currentlabel}{\thepage}}}}}\@gtempa \if@nobreak
    \ifvmode\nobreak\fi\fi\fi\@esphack} \gdef\@eqnlabel{#1}}
    \def\@eqnlabel{}
\def\@vacuum{}
\def\draftmarginnote#1{\marginpar{\raggedright\scriptsize\tt#1}}

\def\draft{
  \oddsidemargin -.5truein
  \def\@oddfoot{\footnotesize \sl preliminary draft \hfil
    \rm\thepage\hfil\sl\today\quad\militarytime}
  \let\@evenfoot\@oddfoot \overfullrule 3pt
    \let\label=\draftlabel
    \let\marginnote=\draftmarginnote
  \def\@eqnnum{(\theequation)\rlap{\kern\marginparsep\tt\@eqnlabel}%
    \global\let\@eqnlabel\@vacuum}

  }

\makeatletter
\newdimen\normalarrayskip              
\newdimen\minarrayskip                 
\normalarrayskip\baselineskip
\minarrayskip\jot
\newif\ifold             \oldtrue            \def\new{\oldfalse}
\def\arraymode{\ifold\relax\else\displaystyle\fi} 
\def\eqnumphantom{\phantom{(\theequation)}}     
\def\@arrayskip{\ifold\baselineskip\z@\lineskip\z@
     \else
     \baselineskip\minarrayskip\lineskip2\minarrayskip\fi}
\def\@arrayclassz{\ifcase \@lastchclass \@acolampacol \or
\@ampacol \or \or \or \@addamp \or
   \@acolampacol \or \@firstampfalse \@acol \fi
\edef\@preamble{\@preamble
  \ifcase \@chnum
     \hfil$\relax\arraymode\@sharp$\hfil
     \or $\relax\arraymode\@sharp$\hfil
     \or \hfil$\relax\arraymode\@sharp$\fi}}
\def\@array[#1]#2{\setbox\@arstrutbox=\hbox{\vrule
     height\arraystretch \ht\strutbox
     depth\arraystretch \dp\strutbox
     width\z@}\@mkpream{#2}\edef\@preamble{\halign
\noexpand\@halignto
\bgroup \tabskip\z@ \@arstrut \@preamble \tabskip\z@ \cr}%
\let\@startpbox\@@startpbox \let\@endpbox\@@endpbox
  \if #1t\vtop \else \if#1b\vbox \else \vcenter \fi\fi
  \bgroup \let\par\relax
  \let\@sharp##\let\protect\relax
  \@arrayskip\@preamble}
\def\eqnarray{\stepcounter{equation}%
              \let\@currentlabel=\theequation
              \global\@eqnswtrue
              \global\@eqcnt\z@
              \tabskip\@centering
              \let\\=\@eqncr
 \halign to \displaywidth\bgroup
    \eqnumphantom\@eqnsel\hskip\@centering
    $\displaystyle \tabskip\z@ {##}$%
    \global\@eqcnt\@ne \hskip 2\arraycolsep
         $\displaystyle\arraymode{##}$\hfil
    \global\@eqcnt\tw@ \hskip 2\arraycolsep
         $\displaystyle\tabskip\z@{##}$\hfil
         \tabskip\@centering
    &{##}\tabskip\z@\cr}
\begingroup\ifx\undefined\newsymbol \else\def\input#1 {\endgroup}\fi
\newfont{\hr}{msbm10}
\newfont{\ams}{msam10}

\voffset= - 1.2in
\hoffset= - 1.0in         

\def\beq{\begin{equation}}
\def\eeq{\end{equation}}
\def\ba{\beq\new\begin{array}{c}}
\def\ea{\end{array}\eeq}
\def\be{\ba}
\def\ee{\ea}
\def\stackreb#1#2{\mathrel{\mathop{#2}\limits_{#1}}}
\def\res{{\rm res}}
\def\Im{{\rm Im}}
\def\F{{\cal F}}
\def\f{{\sf F}}
\def\d{\partial}
\def\N2{${\cal N}=2$}

\def\1N{${\cal N}=1$}
\def\4N{${\cal N}=4$}

\begin{document}
\begin{flushright}
FIAN/TD-02/02\\
ITEP/TH-04/02\\
hep-th/0201267
\end{flushright}
\vspace{0.3cm}

\begin{center}
{\LARGE \bf On Associativity Equations
\footnote{Based on lectures given at the workshops {\em Dualities and Bihamiltonian
Structures in Field and String Theories}, October 2001, SISSA, Italy and {\em Integrable
Models, Strings and Quantum Gravity}, January 2002, Chennai \& Allahabad, India}}\\
\vspace{0.5cm}
{\large A.~Marshakov}\\
\medskip
{\em Theory Department, Lebedev Physics Institute, and\\
Institute of Theoretical and Experimental Physics, Moscow, Russia}\\
{\sf e-mail:\ mars@lpi.ru, andrei@heron.itep.ru}\\
\bigskip\bigskip\medskip

\end{center}

\begin{quotation}
\noindent
We consider the associativity or Witten-Dijkgraaf-Verlinde-Verlinde
(WDVV) equations and discuss one of the most relevant for non-perturbative
physics class of their solutions based on existence of the residue formulas.
It is demonstrated for this case that the proof of associativity equations
is reduced to the problem of solving system of algebraic linear equations.
The particular examples of solutions related to Landau-Ginzburg
topological theories, Seiberg-Witten theories and tau-functions of
quasiclassical hierarchies are discussed in detail. We also discuss
related questions including covariance of associativity equations, their
relation to dispersionless Hirota relations and auxiliary linear problem for
the WDVV equations.
\end{quotation}
\tableofcontents

\section{Introduction
}
\setcounter{footnote}{0}


Integrable systems play now important role in non-perturbative physics.
Beyond the scope of traditional quantum field theory and perturbative string
theory integrable equations
appear to be one of the most effective methods to study the structure of
effective actions, containing information about the exact correlation
functions. Even when the origin of this integrable structure is not yet
understood from the first principles, the observation that some exact
nonperturbative quantities or their generating functions satisfy (systems of)
integrable equations usually reflects the underlying hidden geometric structure,
which, in its turn maybe relevant for better understanding of this object,
from general perspectives of modern string theory.

This paper is devoted to one of the most interesting examples of
nonlinear integrable equations arising in this context~-- the associativity or
WDVV equations \cite{WDVV}, whose appearance
in the framework of topological string theory and nonperturbative
supersymmetric gauge theories is not yet finally understood. Nevertheless,
the scope of application of these equations seem to grow permanently during
last years and it seems that at present one can already see some unique
picture, covering all different and at first glance unrelated examples of
appearance of WDVV equations in modern physics. In what follows we will try
to present first this general picture and then consider different more and
less known examples.

Let us start with basic formulations.
Consider function ${\cal F}$, defined (locally) as a function
of (generally complex) variables ${\bf t}\equiv
\{ t_1,t_2,t_3,\ldots\}$ to be called as {\em times};
dependently on the context, their number can be both finite or infinite.
Sometimes, these variables, when
identified with periods of certain meromorphic differential forms on smooth
Riemann surfaces, for historical reasons will be also denoted as
${\bf a}\equiv\{ a_i\}$.
The associativity or WDVV equations \cite{WDVV} can be in the most
general form written as \cite{MMM}
\be
\label{WDVV}
{\sf F}_i {\sf F}^{-1}_{j}{\sf F}_{k}=
{\sf F}_k {\sf F}^{-1}_{j}{\sf F}_{i}
\;\;\;\;\;\;\mbox{for all}\;\; i,j,k.
\ee
in terms of the matrices
${\sf F}_{i}$ whose matrix elements
\be
\label{matrF}
\|{\sf F}_{i}\|_{jk}=
{\d^3\F\over\d t_i\,\d t_j\,\d t_k} \equiv\F_{ijk}
\ee
are identified with the third derivatives of $\F({\bf t})$.
It will be always assumed below that these matrices and their
generic linear combinations are invertible.
Let us also denote below the second derivatives of $\F$ as
\be
\label{eq:2}
      \F_{ij} \equiv \frac {\d^2 {\cal F}}{\d t_i \d t_j}
\ee
for brevity and make immediately few important remarks.

\begin{itemize}

\item Any function $\F(t)$ of a single variable $t\equiv t_1$ and any
function of two variables $\F(t_1,t_2)$ solves (\ref{WDVV}). In other
words, for functions of one or two variables the WDVV equations
(\ref{WDVV}) are empty. Indeed, in the case of a single variable
one always has $i=j=k$, while in the case of two variables necessarily
either $j=i$ or $j=k$. However, for more than two variables the
associativity equations (\ref{WDVV}) are very nontrivial.

\item  Equations (\ref{WDVV}) are obviously covariant with respect to
linear change of
variables and linear transformations $\f_i \to \sum_k A_{ik}\f_k$. In
particular it means that they can be written in the form
\be
\label{WDVVeta}
{\sf F}_i \eta^{-1}{\sf F}_{k}=
{\sf F}_k \eta^{-1}{\sf F}_{i}
\ee
where for the role of matrix $\eta$
\be
\label{etat}
\eta = \sum \eta_l({\bf t})\f_l
\ee
one can take {\em any} invertible linear combination of $\f_i$ with
generally time-dependent coefficients $\eta_l({\bf t})$. Introducing
so called structure constants
\be
\label{ceta}
{\sf C}_{i}(\eta)=\eta^{-1}{\sf F}_{i}
\ \ \ \ \ \
{\rm or}\\
C_{ij}^{k}(\eta)=\sum_l (\eta^{-1})^{kl}{\cal F}_{ijl}
\ee
one may also rewrite (\ref{WDVV}) as
\be
\label{WDVVeta1}
[{\sf C}_{i}(\eta),{\sf C}_{k}(\eta)]=0
\ee

\item
Covariance under linear transformations imply, in particular, that
it is enough to require existence of the WDVV equations at least for some
fixed $j=j_0$ and $\forall i,k$. For example, associativity equations have
appeared originally in \cite{WDVV} only with fixed $j_0=1$ and
$\f_1=\eta^{(1)}$ called as "metric". However, even
if we know only that eqs. (\ref{WDVV}) hold with some fixed $j_0$,
one may nevertheless transform them into general form
(\ref{WDVVeta}), choosing {\em any} other index to define
"metric" $\eta^{(j)}={\sf F}_j$ (as long as it is non-degenerate) and
structure constants (\ref{ceta}), or restore the generic form of WDVV
equations (\ref{WDVV}). Indeed, since
$\f_i = \eta^{(j_0)}C^{(0)}$, one gets
\be
{\f}_i{\f}_j^{-1}{\f}_k =
\eta^{(j_0)}{\sf C}_i^{(0)}\left(\eta^{(j_0)}{\sf C}_j^{(0)}\right)^{-1}
\eta^{(j_0)}{\sf C}_k^{(0)} =
\eta^{(j_0)}\left({\sf C}_i^{(0)}\left({\sf C}_j^{(0)}\right)^{-1}
{\sf C}_k^{(0)}\right)
\ee
and the r.h.s.
is obviously symmetric w.r.to the permutation $i\leftrightarrow k$ implied
by
\be
[{\sf C}_i^{(0)},{\sf C}_k^{(0)}] = 0, \ \ \ \ \ \ \ \forall i,k
\ee

\item
One should also point out that WDVV equations are {\em not} covariant w.r.t.
arbitrary changes of time-variables since derivatives in (\ref{matrF}) and
(\ref{eq:2}) are not covariant derivatives. We will come back to the problem
of covariance of equations (\ref{WDVV}) \cite{dWM} in sect.~\ref{ss:wdvvdual}.

\item
Sometimes it is convenient to rewrite (\ref{ceta}) as
\be
\label{fceta}
\F_{ijk} = \sum_l C_{ij}^l\eta_{kl}
\ee
and treat as a derivative of certain relation for the {\em second}
derivatives (\ref{eq:2}). Indeed,
choose one of the times, say $t_1$,
and put $\eta_l({\bf t})=\delta_{l1}$ assuming that matrix $\f_1$
is non-degenerate. Then one can
pass from the set of  variables $\{ t_i\}$ to the set of variables $\{{\cal
F}_{j1}\}$ and define the matrix linearly connecting
$ \F_{ijk}$ and $ \eta_{ij}=\F_{ij1}$
\be
\label{cf}
\F_{ijk} = {\d\F_{ij}\over\d t_k} = \sum_l
{\d{\cal F}_{ij}\over\d{\cal F}_{l1}}{\d\F_{l1}\over\d t_k} =
\sum_l C_{ij}^l\F_{kl1}
\ee
as
\be
\label{cff}
C^l_{ij}={\d{\cal F}_{ij}\over\d{\cal F}_{l1}}
\ee
Moreover, the last equality can be understood as a definition of
$C_{ij}^{l}$ even when matrix $\eta$ (\ref{etat}) is not invertible
and will essentially be used, for example, in the infinite-dimensional
situation.

\item Equivalently, the WDVV equations (\ref{WDVV}) can be rewritten as
\be
\label{WDVV2}
\sum_l C_{ij}^l{\cal F}_{lkn} = \sum_l C_{ik}^l{\cal F}_{ljn}.
\ee
or, in other words
\be
\label{x}
X_{ijkn}\equiv\sum_l C_{ij}^l{\cal F}_{lkn}
\ee
is symmetric with respect to permutations of any indices.
Eqs. (\ref{WDVV}), or (\ref{WDVV2}) with $C_{ij}^{l}$  defined
via (\ref{cf}) and (\ref{cff}), are rather restrictive since they
can be considered as an overdetermined set of non-linear differential
equations for the function ${\cal F}$ expressed in the form of relations
for its third order derivatives.

\end{itemize}

These are the most general properties of the equations (\ref{WDVV}) which
do not depend at all of a particular solution. Below, let us first consider some
generalities about the wide class of solutions to (\ref{WDVV}) based on
existence of residue formulas. In sect.~\ref{ss:wdvvgen} we discuss the
residue formulas and prove, that if residue formulas exist the proof of eqs.
(\ref{WDVV}) is reduced to finding solution to a system of algebraic
linear equations.


\section{WDVV equations from associative algebra and residue formulas
\label{ss:wdvvgen}}

\subsection{Associative algebras and residue formulas}

The WDVV equations originally arose \cite{WDVV} as consequence of the
crossing relations
\be
\label{crossing}
\sum_k C^k_{ij}C^n_{kl} = \sum_k C^k_{il}C^n_{kj}
\ee
for the structure constants of the operator algebra of
primary or vacuum operators in two-dimensional topological theories
\be
\label{alg}
\Phi_i\cdot\Phi_j =\sum_k  C^k_{ij}\Phi_k
\ee
where $\{ \Phi_i\}$ are represented by some operators acting in
usually finite-dimensional
Hilbert space of topological theory. It means, in particular, that
in the context of two-dimensional topological theories
(or Seiberg-Witten theories,
see sect.~\ref{ss:sw} below)
one is commonly interested in solutions to WDVV with finite
number of variables, {\it i.e.} the sum in~(\ref{WDVV2}) is finite, though
in principle nothing forbids "infinite-dimensional" topological theories.
Equations (\ref{crossing}) are
just algebraic relations and they turn into the system of nonlinear
differential equations (\ref{WDVV}) only upon
identification of three-point functions of the operators
$\{ \Phi_i \}$ with the third derivatives of some generating
function $\F({\bf t})$
\be
\label{3point}
\langle \Phi_i\Phi_j\Phi_k \rangle = {\d^3\F\over\d t_i\,\d t_j\,\d t_k}
\ee
Indeed, using (\ref{alg}) and (\ref{3point}) one may formally write
\be
\label{alg3p}
\langle \Phi_i\Phi_j\Phi_k \rangle = \sum_k  C^l_{ij}
\langle \Phi_1\Phi_l\Phi_k \rangle
\ee
introducing explicitly the identity operator $\Phi_1\equiv{\bf 1}$, since
only three-point correlators are defined on two-dimensional
world-sheets of simplest spherical topology.
Interpreting 3-point correlator in the r.h.s. of
(\ref{alg3p}) as distinguished "metric" $\eta^{(1)}=\f_1$ one comes to
relation (\ref{fceta})
in its special case (\ref{cf}) which, together with (\ref{crossing}) gives
rise to (\ref{WDVV}).

The presented above "heuristic derivation" and fixing the
distinguished "identity" or puncture operator $\Phi_1\equiv{\bf 1}$
suggests that
two additional requirements on solutions may be imposed:
the constancy of distinguished ``metric'', {\it i.e.} requirement that
matrix $\eta_{ij}=\|\f_1\|_{ij}$ does not depend upon $\{ t_k\}$
and some quasihomogeneity condition on function ${\cal F}$.
Sometimes these requirements are even included into the definition of
the notion of the "WDVV system"~\cite{Dub}, but in what follows we will not
add any extra requirements to the solutions of system (\ref{WDVV})
except for nondegeneracy of matrices in general position.

In topological theories 3-point functions (\ref{3point}) can be often
expressed in terms of so called {\em residue formulas}, which appear as a
consequence of certain "localization" phenomena, typical for these theories.
Up to now there is
no general derivation of residue formulas using localization, but
heuristically the main principle can be illustrated on the basic example of
the Landau-Ginzburg models (to be considered in detail in sect.~\ref{ss:lg}
below), when finite-dimensional algebra (\ref{alg}) may be realized as ring of
all polynomials modulo polynomials vanishing at critical points of some
superpotential $dW(\lambda_\alpha)=0$, and formula (\ref{3point}) also
"localizes" to these critical points, acquiring the form of residue formula
\cite{WDVV,KriW,Dub}
\be
\label{res}
{\cal F}_{ijk} =
\sum_\alpha {\phi_i(\lambda_\alpha)\phi_j(\lambda_\alpha)
\phi_k(\lambda_\alpha)
\over W''(\lambda_\alpha)Q'(\lambda_\alpha)} =
\sum_\alpha \res_{\lambda_\alpha}{\phi_i(\lambda )
\phi_j(\lambda)
\phi_k(\lambda)\over W'(\lambda)Q'(\lambda)} = \\
= \sum_\alpha \res_{\lambda_\alpha}{dH_idH_jdH_k\over
dWdQ} = \oint_{dW=0}{dH_idH_jdH_k\over dWdQ}
\ee
where $dQ(\lambda_\alpha)\neq 0$ and the sum over
residues in the r.h.s. of (\ref{res}) is taken only over part of the poles
(otherwise it would be zero). Operators (\ref{alg}) $\Phi_i$ in (\ref{res})
are represented by polynomials $\phi_i(\lambda)\equiv{dH_i\over d\lambda}$.
For simplicity we consider the case when superpotential is function of
a single variable and, therefore, one
may denote $W'(\lambda)={dW\over d\lambda}$
and $Q'(\lambda)={dQ\over d\lambda}$. However, the properties
of the formula (\ref{res})
to be used below look like not really depending on the number of
variables, and in the case of a single variable (practically
considered everywhere in the paper) universality of formula (\ref{res})
goes far beyond the example of Landau-Ginzburg models.

Formula (\ref{res}) is the main formula to be used throughout the
paper. It the most general situation it was postulated by I.Krichever
\cite{KriW}. In this section we assume that the residue formula is valid
for some function $\F$ and then use this fact to prove that this function
solves the WDVV equations (\ref{WDVV}), demonstrating in particular that
{\em nothing} else should be added except for some "matching" condition
and non-degeneracy, this is one of the main messages of this paper.
Then we consider several examples where residue formula (\ref{res}) can be
established by different methods.

It is necessary to stress that both algebra (\ref{alg}) and residue
formula (\ref{res}) are necessary for the validity of the WDVV equations
(\ref{WDVV}). In sect.~\ref{ss:finite} we demonstrate that
these two ingredients are actually {\em enough} to prove (\ref{WDVV})
and almost nothing extra (except for mentioned above matching and
nondegeneracy) should be imposed. In various
models algebra (\ref{alg}) can be realized in different ways, but for
the class of solutions where residue formula (\ref{res}) exists
it can be
{\em always} presented as algebra of functions (of course, not necessarily
polynomials) modulo functions, vanishing on some submanifold. If this
submanifold as realized as a divisor or set of zeroes of some meromorphic differential
$dW=0$ this algebra is finite-dimensional and leads together
with residue formula (\ref{res}) (certainly with the same $dW$ in
denominator) to the finite system of
differential equations on function $\F$ of finite number of variables.
In sect.~\ref{ss:finite} it is demonstrated that when residue formula
(\ref{res}) is valid the proof of validity of the WDVV equations is reduced to
the problem of solving the system of ordinary (algebraic) linear equations
and for that all extra ingredients of the procedure of constructing solutions to
associativity equations, common in the context of two-dimensional
topological theories \cite{Dub} (constancy of "metric", existence of unity
operator, Frobenius or Darbough-Egoroff structures {\it etc}) are absolutely
inessential.

\subsection{WDVV equations as solving system of linear equations}
\label{ss:finite}

Forget about specifics of the Landau-Ginzburg case and consider formula
(\ref{res}) for a function $\F({\bf t})$ in maximally general setting
\cite{Msusy01}, {\it i.e.}
require now the differential $dW(\lambda)$ in (\ref{res}) to be just
a {\em meromorphic} differential (for example, on arbitrary Riemann surface)
with finite number of zeroes at some points $\{\lambda_\alpha\}$, to be
characterized by values of some adequate for this purpose
co-ordinate $\lambda$, {\it i.e.}
\be
W'(\lambda_\alpha) \equiv {dW\over d\lambda} = 0
\ee
and $Q'(\lambda_\alpha)\neq 0$. In order to get, that such function $\F$
satisfies the system of equations (\ref{WDVV}),
the only extra condition to be imposed is
that matrix $\| \phi_i(\lambda_\beta)\| $ is non-degenerate, {\it i.e.}
\be
\label{det}
\det_{i\alpha}\| \phi_{i}(\lambda_\alpha)\| \neq 0
\ee
In particular, (\ref{det}) requires "matching" $\#(i)=\#(\alpha)$, {\it i.e.} the
number of ``hamiltonians'' $\{ dH_i\}$ or ``fields'' $\{ \phi_i\}$ should be
{\em exactly} equal to the number of zeroes  $\{ \lambda_\alpha \}$ of the
differential $dW$.
One may define now the structure constants $C_{ij}^k$ of the corresponding
finite-dimensional algebra (\ref{alg}) (where the sum is finite) from the
system of {\em linear equations}
\be
\label{eqc}
\phi_i(\lambda_\alpha)\phi_j(\lambda_\alpha) =\sum_k
C^k_{ij}\phi_k(\lambda_\alpha), \ \ \ \ \ \ \ \forall\ \lambda_\alpha
\ee
which hold for {\em all} zeroes $\{ \lambda_\alpha\}$ of $dW$.
Formula (\ref{eqc}) gives a realization of the
finite-dimensional associative algebra (\ref{alg}) defined by any
meromorphic differential $dW$. Using matching and nondegeneracy
conditions (\ref{det}), one can simply solve the system (\ref{eqc}) and write
\be
\label{litc}
C^k_{ij} = \sum_\alpha
\phi_i(\lambda_\alpha)\phi_j(\lambda_\alpha)
\left(\phi_k(\lambda_\alpha)\right)^{-1}
\ee
where the last factor means matrix inverse to $\| \phi_i(\lambda_\alpha)\| $.

The above logic does not change at all, if instead of (\ref{eqc}) we
consider an {\em isomorphic} algebra
\footnote{The situation here is very similar to considered in
\cite{MMM,MMMlong} in the context of algebra of 1-differentials
on Riemann surfaces. However, in
contrast to algebra of forms, algebra of functions (\ref{eqcxi})
is {\em always} associative.}
\be
\label{eqcxi}
\phi_i(\lambda_\alpha)\phi_j(\lambda_\alpha) =\sum_k
C^k_{ij}(\xi)\phi_k(\lambda_\alpha)\cdot\xi(\lambda_\alpha),
\ \ \ \ \ \ \ \forall\ \lambda_\alpha
\ee
with the only requirement $\xi(\lambda_\alpha)\neq 0$, for $\forall\alpha$,
from this point of view (\ref{eqc}) becomes just a particular case of
more general formula (\ref{eqcxi}) with
$\xi(\lambda)\equiv 1$. Then, instead of (\ref{litc}), one immediately gets
\be
\label{litcxi}
C^k_{ij}(\xi) = \sum_\alpha
{\phi_i(\lambda_\alpha)\phi_j(\lambda_\alpha)\over\xi(\lambda_\alpha)}
\left(\phi_k(\lambda_\alpha)\right)^{-1}
\ee
In order to understand when algebra (\ref{eqcxi}) leads to WDVV {\em equations}
(\ref{WDVV}) one should just check consistency between the formulas
(\ref{litcxi}) and (\ref{res}), which can be presented the form
\be
\label{feta}
{\cal F}_{ijk} = \sum_l C_{ij}^l(\xi)\eta_{kl}(\xi)
\ee
with ``metric'' $\eta_{kl}(\xi)$ (which
depends upon $\xi $ in order to cancel dependence of the structure constants)
is non degenerate and satisfies relation (\ref{etat}) or
\be
\label{metric}
\eta_{kl}(\xi) = \sum_a \xi_a {\cal F}_{kla}
\ee
where the third derivatives ${\cal F}_{kla}$ (\ref{3point})
are given by residue formula
(\ref{res}) and $\{\eta_a({\bf t}\}=\{\xi_a\}$ are some coefficients
(which can even depend on times), to be defined below.
Substituting (\ref{res}) into (\ref{metric}) one gets
\be
\label{eta}
\eta_{kl}(\xi) =
\sum_\alpha \res_{\lambda_\alpha}{\phi_k(\lambda)\phi_l(\lambda)\xi(\lambda)
\over
W'(\lambda)Q'(\lambda)} =
\sum_\alpha{\phi_k(\lambda_\alpha)\phi_l(\lambda_\alpha)\xi(\lambda_\alpha)\over
W''(\lambda_\alpha)Q'(\lambda_\alpha)}
\ee
where
\be
\label{xifun}
\xi(\lambda ) = \sum_a\xi_a\phi_a(\lambda )
\ee
Since we already required $\xi(\lambda )$ not to have
zeros in the points $\{\lambda_\alpha\}$, using
condition (\ref{det}) one can always find from (\ref{xifun})
the corresponding coefficients
\be
\label{xia}
\xi_a = \sum_\alpha \xi(\lambda_\alpha )\left(\phi_a(\lambda_\alpha
)\right)^{-1}
\ee
solving again the system of linear equations.

The rest is simple matrix algebra, requiring again {\em only} matching
condition $\#(\alpha)=\#(i)$. Write
\be
\sum_k C_{ij}^k(\xi)\eta_{kl}(\xi) =
\sum_{k,\alpha,\beta}
{\phi_i(\lambda_\alpha)\phi_j(\lambda_\alpha)\over\xi(\lambda_\alpha)}
\cdot\left(\phi_k(\lambda_\alpha)\right)^{-1}\cdot\phi_k(\lambda_\beta)\cdot
{\phi_l(\lambda_\beta)\xi(\lambda_\beta)\over
W''(\lambda_\beta)Q'(\lambda_\beta)}
\ee
and consider it as a product of four matrices. Two mutually inverse
factors in the middle cancel each other and one finally gets
\be
\sum_k C_{ij}^k(\xi)\eta_{kl}(\xi) = \sum_\alpha
{\phi_i(\lambda_\alpha)\phi_j(\lambda_\alpha)\over\xi(\lambda_\alpha)}
{\phi_l(\lambda_\alpha)\xi(\lambda_\alpha)\over
W''(\lambda_\alpha)Q'(\lambda_\alpha)} =
\sum_\alpha
{\phi_i(\lambda_\alpha)\phi_j(\lambda_\alpha)\phi_l(\lambda_\alpha)\over
W''(\lambda_\alpha)Q'(\lambda_\alpha)} = {\cal F}_{ijl}
\ee
and it means that algebra (\ref{eqcxi}) leads to the WDVV equations
(\ref{WDVV}).
Note that derivation is valid for {\em any} function $\xi(\lambda )$ with
the only
restriction that $\xi(\lambda_\alpha)\neq 0$ and, thus, additional
requirements like constancy of "metric" are
absolutely inessential. When all time dependence is
hidden into differential $dW$ the matching condition is satisfied
automatically, at least if $W(\lambda)$ is a polynomial
(the Landau-Ginzburg case, see sect.~\ref{ss:lg}).
Below we also consider two other examples
when, in contrast to the Landau-Ginzburg case, the matching condition is
violated into one or another direction. In the first case
(e.g. Seiberg-Witten prepotential for softly broken ${\cal N}=4$ Yang-Mills
theory, see sect.~\ref{ss:consisw}) one should necessarily add extra variables
to the Seiberg-Witten
periods, in the second case (one of the examples is given by tau-functions
of conformal maps \cite{BMRWZ}, see sect.~\ref{ss:toda}) the situation
is even more striking: $\F =
\log\tau$ satisfies the WDVV equations as a function of only {\em part} of
its variables, when the rest of the variables is fixed.


\section{WDVV equations in topological theories
\label{ss:LG}}

\subsection{Axiomatic of topological theories}

Two-dimensional topological theories may be thought of as string models
or world-sheet theories with almost no excitations
or, better to say, only with (usually finite number of) "vacuum states".
For example in
\N2 SUSY Landau-Ginzburg models, which are defined by
superpotential $W$, these ``vacua'' are identified with the
critical points of the superpotential $dW=0$.
The best way to construct topological theories is to start
with superconformal models with the two-dimensional world-sheet
twisted \N2 supersymmetry and {\em net result} of this
construction, we will be only interested in, can be formulated in
the following axiomatic terms.

\begin{itemize}

\item There is a set (usually finite) of so-called primary fields
$\Phi_{i}$ and corresponding ("vacuum") states $|i\rangle =
\Phi_i|0\rangle$.

\item Among these states there is usually a "trivial" -- or "neutral"
vacuum $|0\rangle$ and corresponding operator is identity $\Phi_1 = {\bf 1}$
or puncture -- in the sense there is "nothing" in corresponding "marked"
point on the world sheet.

\item All ``correlation functions''
can be formulated in terms of three-point functions (\ref{3point}) among
which is a "propagator"
\be
\label{eta-prop}
\eta_{ij} = "\langle\Phi_{i}\Phi_{j}\rangle"
\equiv \langle\Phi_{i}\Phi_{j}{\bf 1}\rangle
\ee

\item Formula (\ref{eta-prop}) identifies naive 2-point function
$\eta_{ij}$ through the 3-point function
$\langle\Phi_{i}\Phi_{j}{\bf 1}\rangle$ with extra puncture operator.
As we already mentioned,
the reason is that it is interpreted from string theory point of view as
a correlation function on sphere -- world-sheet of simplest possible topology,
and there are no well-defined correlation functions on sphere with less than
three marked points.

\end{itemize}

Accepting the above axiomatic definition of topological theory one
finds that all
information is governed by a single function of finite number of
time-variables $\F({\bf t})$ which is usually called a prepotential. In the
case of two-dimensional topological models this function additionally satisfies
\be
\label{eta-const}
\eta_{ij} = \d_{i}\d_{j}\d_1\F({\bf t}) = {\rm const}_{ij}
\ee
and it is in this sense when the first time is called
``distinguished''. However, we will below see that nothing in the discussion
of WDVV equations really depends on condition (\ref{eta-const}).
The main requirement on function $\F({\bf t})$ arises when one considers
the operator algebra (\ref{alg}), which has a structure of
the {\em commutative ring}, with the structure constants
$C_{ij}^{k} = C_{ij}^{k}({\bf t}) =
C_{ji}^{k}({\bf t})$. Applying (\ref{alg}) to the four-point function
$\langle\Phi_{i}\Phi_{j}\Phi_{n}\Phi_{m}\rangle$ and assuming typical in
string theory crossing symmetry
\be\label{wdvv1}
\langle\Phi_{i}\Phi_{j}\Phi_{n}\Phi_{m}\rangle
= C_{ij}^{k}
\langle\Phi_{k}\Phi_{n}\Phi_{m}\rangle =
C_{in}^{k}
\langle\Phi_{k}\Phi_{j}\Phi_{m}\rangle
\ee
one gets the relations (\ref{crossing})
(where indices are raised with the help of propagator $\eta_{ij}$,
playing the role of "metric"), giving rise to WDVV equations (\ref{WDVV}).
Hence, the solutions to WDVV equations valid for two-dimensional
topological theories should satisfy additional requirement
(\ref{eta-const}), such solutions were classified in \cite{Dub}.

The integrable structures underlying topological two-dimensional models can be
elegantly described by the following zero-curvature condition
\cite{CeVa,Dub}
\be\label{dubrovin}
[\nabla _{i},\nabla _{j}] = 0
 \\
\nabla _{i} = {\sf 1}\cdot{\d\over\d t_{i}} - \zeta{\sf C}_{i}
\ee
where $\|{\sf 1}\|_j^k=\delta_j^k$ is unit matrix and
$\|{\sf C}_{i}\|_{j}^{k} = C_{ij}^{k}$.
Due to explicit dependence on the additional
"spectral parameter" $\zeta$ the equation
(\ref{dubrovin}) is equivalent to two independent conditions: quadratic
$\zeta^2$-term gives
\be\label{wdvv3}
[{\sf C}_{i},{\sf C}_{j}] = 0
\ee
or the crossing relations (\ref{crossing}), while linear in $\zeta$ term leads
to
\be
\label{dc}
\zeta\left(\d_{i}C_{jk}^{l}-\d_{j}
C_{ik}^{l}\right) = 0
\ee
Using (\ref{ceta}), (\ref{fceta}) and the fact that the "metric" $\eta$
(\ref{eta-const}) is constant the formula (\ref{dc}) can be rewritten as
\be
\label{curvf}
\d _{[j}\F_{i ]kl} = 0
\ee
what means (at least locally) that $\F_{ijk} =
\d_{i}\F_{jk}$ or taking into account the fact that
$\F_{ijk}$ are symmetric in all indices, relation (\ref{curvf}) can be
integrated up to
\be
\F_{ij}={\d^2\F\over\d t_i\d t_j}
\\
\F_{ijk} = {\d^3\F\over\d t_i\d t_j\d t_k}
\ee
{\it i.e.} from (\ref{curvf}) one gets that 3-point functions (\ref{3point})
$\F_{ijk}$ related to structure constants of topological theory are third
derivatives of some function $\F$. It means that WDVV equations in
topological theories (with constant "metric") are indeed encoded in
(\ref{dubrovin}). We will come back to discussion of auxiliary linear problem
(\ref{dubrovin}) in sect.~\ref{ss:auli}.

\subsection{Landau-Ginzburg models
\label{ss:lg}}

These are the
topological theories determined by polynomial
superpotential in general of several complex variables. In the simplest
situation the superpotential is just a polynomial $W(\lambda )$ of complex
variable $\lambda$ and of degree $N$
\be
W(\lambda) = \lambda^N + \sum_{k=0}^{N-2} u_k \lambda^k
\ee
which has $(N-1)$ parameters or ``degrees of freedom''. The primaries
are defined by the formulas
\be
\label{dLG}
\phi_k(\lambda) = {\d W\over \d t_k} = \left({d\over d\lambda}
W^{k/N}\right)_+
\ee
($k=1,\dots,N-1$), where ``+'' means polynomial part in $\lambda$
of the r.h.s. In particular,
\be
\label{d1}
\phi_1(\lambda) = {\d W\over\d t_1} = 1
\ee
corresponds to the $\Phi_1={\bf 1}$ unity operator.
The corresponding times or "flat coordinates" are given by
\be
\label{tkri}
t_k = - {N\over k(N-k)}\res_{\infty}
\left( W^{1-k/N}d\lambda\right)
\ee
It is easy to check that (\ref{tkri}) is indeed consistent with
(\ref{dLG}). Taking derivatives of (\ref{tkri}) one gets
\be
{\d t_k\over\d t_j} = -{1\over k}\res_\infty
\left(W^{-k/N}{\d W\over\d t_j}\right) = -{1\over k}\res_\infty
\left(W^{-k/N} \d_\lambda W^{j/N}_+\right) = -{1\over k}\res_\infty
\left(W^{-k/N} \d_\lambda W^{j/N}\right) = \delta_{jk}
\ee
where we used that $\res_\infty {d\lambda\over\lambda} =
- \res_0 {d\lambda\over\lambda} = -1$.

The primaries (\ref{dLG}) satisfy the associative algebra (\ref{alg})
\be
\label{ringw}
\phi_i(\lambda)\phi_j(\lambda) = \sum_{k=1}^{N-1}
C_{ij}^k\phi_k(\lambda) + R_{ij}(\lambda)W'(\lambda)
\ee
which is nothing but a factor of the ring of
all polynomials over the ideal
$W'(\lambda)=0$ {\it i.e.}
polynomials $\phi_k(\lambda)$ defined in (\ref{dLG}) span
${{\bf C}[\lambda]\over W'(\lambda)}$
\footnote{Of course from the point of view of algebra (\ref{alg}),
(\ref{ringw}) basis (\ref{dLG}) is just one of many possible basises.
It is distinguished only from the point of view of residue formula
(\ref{res}) (see below) which depends crucially upon the choice of the basis
and so do the WDVV equations (\ref{WDVV})}.
It means that the crossing relation
(\ref{crossing}) for commutative ring (\ref{ringw}) is satisfied
automatically.

The proof of the validity of the WDVV equations is, therefore,
essentially the check of
consistency between the formulas (\ref{ringw}) and particular case
of (\ref{res}) with $Q'(\lambda)=1$
\be
{\cal F}_{ijk} = {\rm res}_{W'(\lambda)=0}\frac{\phi_i(\lambda)\phi_j(\lambda)
\phi_k(\lambda)}{W'(\lambda)} \equiv \sum_{\alpha}
\frac{\phi_i\phi_j\phi_k(\lambda_\alpha)}
{W''(\lambda_\alpha)}
\label{vc}
\ee
as well as
\be
\label{etaLG}
\eta_{kl} \equiv \eta^{(LG)}_{kl} =
\sum_\alpha \res_{\lambda_\alpha}{\phi_k(\lambda)\phi_l(\lambda)
\over W'(\lambda)}
\equiv \sum_{\alpha}
\frac{\phi_k\phi_l(\lambda_\alpha)}
{W''(\lambda_\alpha)}
\ee
where $\lambda_\alpha$ are the critical points of the
Landau-Ginzburg superpotential of $W'(\lambda_\alpha)=0$,
and (\ref{fceta}).

In addition to the consistency of eqs.~(\ref{ringw}), (\ref{etaLG})
and (\ref{fceta})
one should remember that ${\cal F}_{ijk}$ given by
(\ref{vc}) are third derivatives of a single function ${\cal F}({\bf t})$.
This function can be defined, for example, saying that its first
derivatives are given by "dual" to (\ref{tkri}) formula
\be
\label{mtkri}
{\d\F\over\d t_k} = {N\over N+k}\res_{\infty}\left(W^{1+{k\over
N}}d\lambda\right)
\ee
It is easy to check then that the second derivatives
\be
\label{2derLG}
\F_{ik} = {\d^2\F\over\d t_i\d t_k} = \res_{\infty}\left(W^{k/N}
{\d W\over\d t_i}\right) = \res_{\infty}\left(W^{k/N}\d_\lambda W^{i/N}_+\right)
\ee
and the expression in r.h.s. of (\ref{2derLG}) is indeed symmetric w.r.t.
$(i\leftrightarrow k)$.

For the third derivatives, using simple algebra, one gets
\be
\label{3derLG}
\F_{ijk} = \res_{\infty}\left({i\over N}W^{i/N-1}{\d W\over\d t_k}
\d_\lambda W^{j/N}_+\right) + \res_{\infty}\left(W^{i/N}
{\d\over\d\lambda}\left({j\over N} W^{j/N-1}{\d W\over\d t_k}\right)_+\right)
= \\ =
\res_{\infty}\left({i\over N}W^{i/N-1}\d_\lambda W^{j/N}_+
\d_\lambda W^{k/N}_+\right) -
\res_{\infty}\left(\d_\lambda W^{i/N}
\left({j\over N} W^{j/N-1}\d_\lambda W^{k/N}_+\right)_+\right)
= \\ =
\res_{\infty}{\d_\lambda W^{i/N}\d_\lambda W^{j/N}_+
\d_\lambda W^{k/N}_+\over W'}\ - \\ -
\res_{\infty}\left(\d_\lambda W^{i/N}
{j\over N} W^{j/N-1}\d_\lambda W^{k/N}_+\right) +
\res_{\infty}\left(\d_\lambda W^{i/N}
\left({j\over N} W^{j/N-1}\d_\lambda W^{k/N}_+\right)_-\right)
= \\ =
\res_{\infty}{\d_\lambda W^{i/N}\d_\lambda W^{j/N}_+
\d_\lambda W^{k/N}_+\over W'}\ - \\ -
\res_{\infty}{\d_\lambda W^{i/N}\d_\lambda W^{j/N}
\d_\lambda W^{k/N}_+\over W'} +
\res_{\infty}{\d_\lambda W^{i/N}_+\d_\lambda W^{j/N}
\d_\lambda W^{k/N}_+\over W'}
= \\ =
\res_{\infty}{\d_\lambda W^{i/N}\d_\lambda W^{j/N}_+
\d_\lambda W^{k/N}_+\over W'}\ -
\res_{\infty}{\d_\lambda W^{i/N}_-\d_\lambda W^{j/N}
\d_\lambda W^{k/N}_+\over W'}
= \\ =
\res_{\infty}{\d_\lambda W^{i/N}_+\d_\lambda W^{j/N}_+
\d_\lambda W^{k/N}_+\over W'}\ -
\res_{\infty}{\d_\lambda W^{i/N}_-\d_\lambda W^{j/N}_-
\d_\lambda W^{k/N}_+\over W'}
= \\ =
\res_{\infty}{\d_\lambda W^{i/N}_+\d_\lambda W^{j/N}_+
\d_\lambda W^{k/N}_+\over W'}
\ee
{\it i.e.} the residue formula (\ref{vc}) (the total sign and/or any numerical
coefficient is inessential)
\be
\F_{ijk} = \res_{\infty}{\phi_i(\lambda)\phi_j(\lambda)\phi_k(\lambda)
\over W'} = - \stackreb{W'=0}{\res}{\phi_i(\lambda)\phi_j(\lambda)\phi_k(\lambda)
\over W'}
\ee
Then the proof of
(\ref{fceta}) is straightforward, since
\be
\sum_l\eta_{kl}C^l_{ij} = \sum_{l,\alpha}
\frac{\phi_k\phi_l(\lambda_\alpha)}
{W''(\lambda_\alpha)} C^l_{ij} \stackrel{(\ref{ringw})}{=} \\ =
\sum_{\alpha}
\frac{\phi_k(\lambda_\alpha)}
{W''(\lambda_\alpha)} \phi_i(\lambda_\alpha)\phi_j(\lambda_\alpha)
= {\cal F}_{ijk}.
\ee
Note that (\ref{ringw}) is defined modulo $W'(\lambda)$,
but $W'(\lambda_\alpha) = 0$ for all $\lambda_\alpha$.

A remark here should be made about "topological" or "Landau-Ginzburg
metric" which is defined by pairing (\ref{etaLG})
and is distinguished in the context of two-dimensional Landau-Ginzburg models
(for example it is constant, as it should be for a topological theory).
Comparing (\ref{etaLG}) and (\ref{eta}) it is
easy to see that the "Landau-Ginzburg metric" is just a particular choice
in (\ref{eta}) $\eta^{(LG)}_{kl} = \eta_{kl}(\xi^{(LG)})$ provided by
special choice
\be
\xi^{(LG)}(\lambda_\alpha) =
\sum_b\xi^{(LG)}_b\phi_b(\lambda_\alpha ) = Q'(\lambda_\alpha), \ \ \ \ \
\forall \lambda_\alpha
\ee
and $\{\xi^{(LG)}_b \}$ can be again easily found under the only
condition (\ref{det}).
Hence, the "Landau-Ginzburg metric" (\ref{etaLG}) arises just as a particular
case of our general consideration.
Of course, nothing guarantees constancy of the "Landau-Ginzburg metric"
(\ref{etaLG}) in general situation, say when gravitational descents are
"switched one" and nontrivial function $Q'(\lambda)\neq 1$ is
determined by "gravitational times". However, for the
"small phase space" when all gravitational times vanish,
$\xi^{(LG)}(\lambda)=Q'(\lambda)=1$ and we are coming
back to (\ref{eqc}), (\ref{litc}), (\ref{vc}) and (\ref{etaLG}).

Note finally, that the ``Gauss-Manin'' relation
(see, for example, \cite{ItoYang})
\be
\label{gm}
\d_\lambda R_{ij} = {\d^2 W\over\d t_i\d t_j}
\ee
where $R_{ij}(\lambda)$ is defined in (\ref{ringw}),
is a trivial consequence of the above formulas for the Landau-Ginzburg
models (valid, of course,
only in the distinguished variables (\ref{tkri})).
Indeed, the sense of algebra (\ref{ringw}) is very simple, it says that
the expression
\be
{\phi_i(\lambda)\phi_j(\lambda)\over W'(\lambda)} - R_{ij}(\lambda)
\ee
is ratio with $W'$ in denominator and numerator being
a polynomial of degree less than $(N-1)$ which therefore can be decomposed
into a linear combination $\sum_{k=1}^{N-1}C_{ij}^k\phi_k$. Using
(\ref{dLG}) one can perform this rather simple exercise. Indeed,
\be
\label{ex}
{\phi_i(\lambda)\phi_j(\lambda)\over W'(\lambda)} =
{\left({d\over d\lambda}W^{i/N}\right)_+
 \left({d\over d\lambda}W^{j/N}\right)_+
  \over W'(\lambda)} =
\\
= { \left({i\over N} W^{i/N-1}W'\right)_+
   \left({j\over N} W^{j/N-1}W'\right)_+ \over W'}
  \equiv {(AW')_+(BW')_+\over W'} \equiv {(AW')_+ R\over W'}
\ee
where we have introduced for simplicity of notations the Laurent series
\be
\label{AB}
A(\lambda) \equiv {i\over N} W^{i/N-1}(\lambda) =
{i\over N}\lambda^{i-N} + \dots
\\
B(\lambda) \equiv {j\over N} W^{j/N-1}(\lambda) =
{j\over N}\lambda^{j-N} + \dots
\ee
and polynomial
\be
\label{R}
R(\lambda) \equiv (BW')_+ = {j\over N}\lambda^{j} + \dots
\ee
It is quite easy to divide two polynomials and get
\be
\label{div}
{(AW')_+ R\over W'} = {AW'R - (AW')_- R\over W'} = AR - {(AW')_- R\over W'}
= \\
= (AR)_+ + \left((AR)_- - {(AW')_- R\over W'}\right) =
(AR)_+ + {W'(AR)_- - (AW')_- R\over W'}
\ee
Now one can see that the numerator of the second term in the last
expression $(W'(AR)_- - (AW')_- R)$ has degree not more than
$(N-2)$ (and therefore corresponds to $\sum_{k=1}^{N-1}C_{ij}^k\phi_k$).
Indeed, this simply follows from looking at the highest degrees in
(\ref{AB}) and (\ref{R}). The first term, thus corresponds to polynomial
part of the result of division (\ref{div}) and gives rise to the relation
\be
\label{Q}
R_{ij}(\lambda) = (AR)_+ = \left({i\over N} W^{i/N-1}
\left({j\over N} W^{i/N-1}W'\right)_+\right)_+
\ee
This is exactly what we need in order to establish relation (\ref{gm})
since from (\ref{dLG}) it is easy to get
\be
\label{d2W}
{\d^2 W\over\d t_i\d t_j} = {d\over d\lambda}{\d\over\d t_j} W^{i/N}_+ =
{d\over d\lambda}\left({i\over N}W^{i/N-1}{\d W\over\d t_j}\right)_+
= \\ =
{d\over d\lambda}\left({i\over N}W^{i/N-1}{d\over d\lambda}
\left(W^{j/N}\right)_+\right)_+ =
{d\over d\lambda}\left({i\over N}W^{i/N-1}
\left({j\over N}W^{j/N-1}W'\right)_+\right)_+
\ee
Comparison of (\ref{Q}) and (\ref{d2W}) gives (\ref{gm}). This relation will
be used below in sect.~\ref{ss:auli}.


\section{Associativity equations in Seiberg-Witten theory
\label{ss:sw}}

\subsection{Seiberg-Witten theory: period matrices and residue formulas
\label{ss:ressw}}

In Seiberg-Witten theory \cite{SW}
the role of function ${\cal F}$ is played by prepotential
defined with the help of family of auxiliary Riemann surfaces,
endowed with some special meromorphic differential $dS$.
It satisfies the system of WDVV equations (\ref{WDVV}) provided the times
$\{ t_i\} $ are identified with the periods
\be\label{aper}
a_j = \oint _{A_j} dS
\ee
generating meromorphic differential \cite{MMM}.
The prepotential ${\cal F}$ is defined implicitly by
\be\label{aad}
a^D_i = {\partial {\cal F}\over\partial a_i}
\ee
where (by accepted convention called ${\bf a}_D$)
\be\label{adper}
a^D_j = \oint _{B_j}dS
\ee
is the set of dual ${\bf B}$-periods.
In particular, in the specific co-ordinates $\{ a_i\} $
the matrix of second derivatives of the function $\F$
\be
\label{prepsw}
{\d^2{\cal F}\over\d a_i\d a_j} = T_{ij}({\bf a})
\ee
plays the role of the
{\em period matrix} of Riemann surface $\Sigma $ (remember, that
these are {\em not} covariant
derivatives and therefore this is co-ordinate dependent statement).
For example, in the case of pure \N2 SUSY gauge
theory with the $SU(N)$ gauge group the auxiliary Riemann surface and
meromorphic differential have the form \cite{sun}
\be
\label{suncu}
w + {\Lambda^{2N}\over w} = P_{N}(\lambda)
\\
dS = \lambda{dw\over w}
\ee
or
\be
\label{c2}
y^2 = P_N(\lambda)^2 - 4\Lambda^{2N}
\\
dS = \lambda{dP_N\over y}
\ee
On genus $g=N-1$ Riemann surface $\Sigma _g$ (\ref{suncu}) there
exists $2g$ independent noncontractable contours
which can be split into so called
${\bf A}\equiv\{ A_i\}$ and ${\bf B}\equiv\{ B_i\}$,
$i = 1,\dots,g$, cycles with the
intersection form $A_i\circ B_j = \delta _{ij}$. Relabeling of $A$- and
$B$-cycles is called a {\em duality} transformation and the covariance of
WDVV equations (\ref{WDVV}) under duality transformations \cite{dWM} will
be discussed in sect.~\ref{ss:wdvvdual}.
The holomorphic differentials
are usually taken to be normalized to the ${\bf A}$-cycles
\be\label{normA}
\oint _{A_j}d\omega _i = \delta _{ij}
\ee
then their integrals along the ${\bf B}$-cycles give the period matrix
\be\label{pemat}
\oint _{B_j}d\omega _i =  T_{ij}
\ee
The period matrix (\ref{pemat}) is symmetric, it can be checked
by direct application of the Stokes theorem to the surface integral
\be\label{symmpemat}
0 = \int _{\Sigma _g}d\omega _i\wedge d\omega _j =
\sum _{k = 1}^g \oint _{A_k}d\omega _i\oint _{B_k}d\omega _j -
(i\leftrightarrow j) = T_{ij} - T_{ji}
\ee
The generating differential (in particular case given by (\ref{suncu}))
should satisfy the basic property
\be
\label{hol}
\delta_{\rm moduli} dS = {\rm holomorphic}
\ee
It is easy to see that the holomorphic differentials in the r.h.s. of
(\ref{hol}) become canonically normalized  (\ref{normA}) if one takes as
co-ordinates on moduli
space the ${\bf A}$-periods of the differential $dS$ (\ref{aper}).
Indeed,
\be
\delta a_j = \oint _{A_j} {\d dS\over\d a_i}\delta a_i
\label{vara}
\ee
and comparing (\ref{vara}) with (\ref{normA}) one gets
\be
{\d dS\over\d a_i}= d\omega_i
\label{varsa}
\ee
where equality means in fact equality modulo total derivatives,
disappearing from the integrals over closed contours.

Taking derivatives of both sides of (\ref{adper})
w.r.t. $a_i$ and using (\ref{varsa}),
(\ref{pemat}) one gets that
\be
{\d a^D_j\over\d a_i} = \oint _{B_j}{\d dS\over\d a_i} =
\oint _{B_j}d\omega_i = T_{ij}
\label{ada}
\ee
In the Seiberg-Witten case there is no distinguished index $i$: all
arguments $a_i$ of prepotential can be treated on equal footing. Thus,
if some kind of WDVV equations holds in this case, it should be invariant under
any permutation of the indices $i,j,k$ -- criterium satisfied by
the system (\ref{WDVV}). Formulas (\ref{ada}) and (\ref{symmpemat}) lead
to (\ref{aad}) and (\ref{prepsw}).

In the case of hyperelliptic Riemann surfaces one may prove the existence of
residue formula using the relation
\cite{Fay} between the derivatives of the matrix elements of the period
matrix w.r.t. positions of the branch points with the values of
{\em canonical} holomorphic differentials at
branch points of hyperelliptic curves.
Indeed, it is easy to check that the derivative of the period
matrix of hyperelliptic curve w.r.t. the variation
of any branch point $\lambda_I$ of the equation (\ref{c2})
can be expressed as \cite{Fay}
\be
\frac{\partial
T_{ij}}{\partial \lambda_I} = \hat
\omega_i(\lambda_I)\hat\omega_j(\lambda_I).
\label{derram}
\ee
This follows from a particular Riemann bilinear identity -- an analog of
relation (\ref{symmpemat})
\be
\label{biri}
0=\int d\omega_i\wedge {\partial d\omega_j\over\partial\lambda_{I}}=
\sum_k\left( \oint_{B_k}{\partial
d\omega_j\over\partial\lambda_{I}}\oint_{A_k}d\omega_i-\oint_{A_k}{\partial
d\omega_j\over\partial\lambda_{I}}\oint_{B_k}d\omega_i \right) -
\stackreb{\lambda_{I}}{\hbox{res}}\left(\omega_i
{\partial d\omega_j\over\partial\lambda_{I}}\right)
\ee
The last term appears since the derivative
of holomorphic differential w.r.t. position of branch point $\lambda_I$
acquires a pole at $\lambda=\lambda_I$.
Let us define the co-ordinates of the branch points $\lambda_I$,
rewriting (\ref{c2}) as a product
over $2N=2g+2$ roots
\be
\label{c2branch}
y^2 = P_N^2(\lambda )-4\Lambda^{2N} =
\prod_{I=1}^{2N}(\lambda - \lambda_I)
\ee
and denote $\hat y^2(\lambda_I) \equiv
\prod_{J\neq I}(\lambda_I - \lambda_J)$. Indeed, in the vicinity of
$\lambda_I$ one may use co-ordinate $\sqrt{\lambda-\lambda_I}$ and the
expansion
\be
d\omega_i= 2\hat \omega_i(\lambda_{I})d\sqrt{\lambda-\lambda_{I}}
+ \dots
\\
{\partial \omega_j\over\partial\lambda_{I}} = - {\hat
\omega_j(\lambda_{I})\over\sqrt{ \lambda-\lambda_{I}}} + \dots
\ee
where
\be
\label{omegahat}
\hat \omega_i(\lambda_{I}) =
{\phi_i(\lambda_I)\over\sqrt{\prod_{J\neq I}
(\lambda_I-\lambda_J)}}
\ee
Hence, bilinear relation (\ref{biri}) together with (\ref{normA}) and
(\ref{pemat}) gives
\be
\label{tres}
{\d T_{ij}\over\d\lambda_I} = \stackreb{\lambda_{I}}{\hbox{res}}\left(\omega_i
{\partial d\omega_j\over\partial\lambda_{I}}\right)
\ee
Using the expansion in the vicinity of the point $\lambda_{I}$,
the computation of residue in (\ref{tres}) gives rise to (\ref{derram}).

However, for the family (\ref{c2branch}) all $2N=2g+2$
branch points $\{\lambda_I\}$ depend only on $g$ moduli $\{u_k\}$ or
$\{a_i\}$. It means that
\be
\label{oola}
\frac{\partial^3 \F}{\partial a_i\partial a_j\partial a_k} =
\frac{\partial T_{ij}}{\partial a_k}
= \sum_I \frac{\partial T_{ij}}{\partial\lambda_I}
\frac{\partial \lambda_I}{\partial a_k} = \\ =
\sum_I \hat\omega_i(\lambda_I)
\hat\omega_j(\lambda_I)
\frac{\partial \lambda_I}{\partial a_k}
\ee
Variation of the equation of the curve (\ref{suncu}) (at fixed $w$) gives
\be
\delta\lambda = - {\sum_k \lambda^k\delta u_k\over P_N'(\lambda)}
\ee
which, taken at $\lambda=\lambda_I$, together with (\ref{phiua}) leads
to
\be
\label{lambdaa}
{\d\lambda_I\over\d a_i} = - {\sum_k \lambda_I^k{\d u_k\over\d
a_i}\over P_N'(\lambda_I)} =
- {\phi_i (\lambda_I)\over P_N'(\lambda_I)} = - {\hat\omega
(\lambda_I) \hat y(\lambda_I)\over P_N'(\lambda_I)}
\ee
Substituting (\ref{lambdaa}) into (\ref{oola}) one finally gets
\be
{\cal F}_{ijk} = \frac{\partial^3{\cal F}}{\partial a_i\partial a_j
\partial a_k} = \frac{\partial T_{ij}}{\partial a_k} =  \\
=\sum_{I} \frac{\hat\omega_i(\lambda_I)\hat\omega_j
(\lambda_I)\hat\omega_k(\lambda_I)}{P'_N(\lambda_I)
/\hat y(\lambda_I)} =
\stackreb{d\lambda =0}{{\rm res}} \frac{d\omega_id\omega_j
d\omega_k}{d\lambda\frac{dP_N}{y}} =
\stackreb{d\lambda =0}{{\rm res}} \frac{d\omega_id\omega_j
d\omega_k}{d\lambda\left(\frac{dw}{w}\right)}
\label{v}
\ee
Notice, that
the sum in this formula runs over all $2N=2g+2$ branch points
$\lambda_I$ of hyperelliptic curve (\ref{c2branch}).
We will return to this fact below in sect.~\ref{ss:consisw}.

Since the integrand in (\ref{v}) is holomorphic one may also rewrite the
residue formula as
\be
\label{v1}
{\cal F}_{ijk} = -
\stackreb{{dw\over w} =0}{{\rm res}} \frac{d\omega_id\omega_j
d\omega_k}{d\lambda\left(\frac{dw}{w}\right)}
\ee
which will be especially important in sect.~\ref{ss:consisw} below.

The hyperelliptic case is distinguished by existence of very convenient
co-ordinates on moduli space -- the branch points $\{ \lambda_I\}$
(\ref{c2branch}).
We do not have any useful analog in generic situation. However,
ideologically situation is quite similar and derivatives w.r.t. moduli
produce new singularities. It means, that if one takes as co-ordinates on
moduli space, say, the coefficients of embedding polynomials, an analog of
the formula (\ref{tres}) should exist in the form (see, for example,
discussion in \cite{MMMlong} of the Calogero-Moser case)
\be
\frac{\partial T_{ij}}{\partial u_s} =
\res \left( \omega_i \frac{\partial d\omega_j}{\partial u_s}\right)
\label{prom1}
\ee
Formulas like (\ref{prom1}) generally give rise to
\be
\frac{\partial T_{ij}}{\partial s^k} =
\stackreb{dz = 0}{\res}
\frac{d\omega_id\omega_jdv_k}{dz d\lambda} \\
\frac{\partial^3 {\cal F}}{\partial a^i\partial a^j\partial a^k} =
\frac{\partial T_{ij}}{\partial a^k} =
\stackreb{dz = 0}{\res}
\frac{d\omega_id\omega_jd\omega_k}{dz d\lambda}
\label{resfor}
\ee
as proposed in \cite{KriW}, but their proof in general form is beyond the
scope of this paper. As we also see below, generally it is not enough
to consider
formulas like (\ref{resfor}) restricted to holomorphic differentials.

\subsection{The associativity equations in Seiberg-Witten theory and
algebras of forms
\label{ss:proof}}

Imagine now that in topological Landau-Ginzburg models
(\ref{ringw}), (\ref{etaLG})
we change the definition of "metric"
\be
\eta^{(LG)}_{kl} \rightarrow \eta_{kl}(\xi) =
\sum_{\alpha}
\frac{\phi_k\phi_l(\lambda_\alpha)}
{W''(\lambda_\alpha)}\xi(\lambda_\alpha).
\ee
which is equivalent to general definition (\ref{metric}), dependent
on some function $\xi(\lambda)$ (\ref{xifun}), which is now required to be a
polynomial. As we already know
from sect.~\ref{ss:finite} nothing is changed for the proof
of the WDVV equations, provided the
algebra (\ref{ringw}) is also changed for (\ref{eqcxi}), or, to
the polynomial algebra
\be
\phi_i(\lambda)\phi_j(\lambda) = C_{ij}^k(\xi)\phi_k(\lambda)
\xi(\lambda)\ {\rm mod}\ W'(\lambda).
\label{algp}
\ee
Eq.~(\ref{algp}) describes an associative polynomial algebra whenever the
polynomials $\xi(\lambda)$ and $W'(\lambda)$ are co-prime
{\it i.e.} do not have common divisors. As in general situation of
sect.~\ref{ss:finite} formula (\ref{vc}) and thus the fact that
${\cal F}_{ijk}$ are third derivatives of the same ${\cal F}$ remains
intact.

In the Seiberg-Witten theory the polynomials
$\phi_i(\lambda)$ (\ref{dLG}) are substituted by canonical holomorphic
differentials $d\omega_i(\lambda )$ on Riemann surface \cite{MMM}
(in hyperelliptic case given, for example, by formula (\ref{c2})).
Instead of (\ref{ringw}) and (\ref{algp}) one may also replace polynomial
$\xi(\lambda)$ by some differential $dG$ and
introduce formally the {\em algebra of forms} \cite{MMMforms,MMMlong}
\be
d\omega_i(\lambda )d\omega_j(\lambda ) =
C_{ij}^k(dG ) d\omega_k(\lambda )
dG (\lambda ) \ {\rm mod}\ \frac{dP_N(\lambda )d\lambda }{y^2}.
\label{algforms}
\ee
In contrast to (\ref{ringw}) and (\ref{algp}) in the algebra of forms
one cannot simply choose $dG = 1$
to reproduce an analog of (\ref{ringw}) since $dG$ is a
1-form, instead, one can just require, for example, $dG $ to be a holomorphic
differential.
There is no distinguished
holomorphic differential among $g$-parametric family, and $dG $
can be {\em any} one from this family with the only requirement that it is
co-prime with $\frac{dw}{w}=\frac{dP_N(\lambda )}{y}$.

If algebra (\ref{algforms}) exists and is associative (which is
nontrivial for the
algebra of forms and is even wrong in generic situation, see
\cite{MMMlong}), the structure constants $C_{ij}^k(\xi )\equiv
C_{ij}^k(dG )$ satisfy the associativity condition (\ref{crossing})
(if $dG $ and
${dP_N\over y}$ are co-prime). It is easy to show that
it indeed exists, {\it i.e.} for given $dG $ one can find
$C_{ij}^k(dG)$ in the case
of hyperelliptic surfaces, since all
canonical holomorphic differentials $d\omega_i$ (\ref{normA}) are
linear combinations of
\be
dv_k(\lambda ) = \frac{\lambda ^{k-1}d\lambda }{y}, \ \ \ k=1,\ldots,g
\ee
{\it i.e.}
\be
\label{holdif}
d\omega_i = {\phi_i(\lambda)d\lambda\over y} = \sum_k (\sigma^{-1})_{ik}dv_k
\ee
where $\{ \phi_i(\lambda)\}$ are certain polynomials of degree not exceeding
$(N-2)$. The coefficients $\sigma_{ki}$ may be defined as
\be
\sigma_{ki} = \oint_{A_i}dv_k = {\d a_i\over\d u_k}
\label{sigmadef}
\ee
where $\{ u_k\}$ are coefficients of the polynomial $P_N(\lambda)$,
in (\ref{suncu}) and (\ref{c2}), the last relation follows
from (\ref{hol}), (\ref{aper}) and (\ref{varsa}).
In other words, for the polynomials $\{\phi_i(\lambda)\}$ related to
canonical holomorphic differentials (\ref{holdif}) one gets
\be
\label{phiua}
\phi_i(\lambda) = \sum _k {\d u_k\over\d a_i}\ \lambda^k
\ee
Thus, algebra of forms (\ref{algforms}) on hyperelliptic surfaces is in fact
isomorphic to the polynomial ring (\ref{algp}) where the role of
generators is played by polynomials $\phi_i = yd\omega_i/d\lambda$
(\ref{holdif}) and the role of (derivative of)
superpotential -- by $W'(\lambda)= P'_N(\lambda) = {y dw/w\over d\lambda}$
\be
\phi_i(\lambda)\phi_j(\lambda)= \sum_k
C_{ij}^k \phi_k(\lambda)\xi(\lambda) +
R_{ij}(\lambda )P'_N(\lambda )
\label{polrcou}
\ee
where we now fix $\xi(\lambda)=ydG/d\lambda$.
In the l.h.s. of eq.~(\ref{polrcou}) one has a polynomial of degree $2(g-1)$
and since $P'_N(\lambda )$ has degree $N-1=g$,
$R_{ij}(\lambda )$ should be of degree
$2(g-1)-g = g-2$. The identification of two polynomials of
degree $2(g-1)$ by equality (\ref{polrcou}) imposes a set of
$2g-1$ equations for the coefficients. One may also say, that
we have freedom to adjust $C_{ij}^k$ and $R_{ij}(\lambda )$
(with $i,j$ fixed), {\it i.e.} $g + (g-1) = 2g-1$ free parameters:
exactly what is necessary for unique solution of the system of linear
equations. The corresponding system of linear equations
is non-degenerate for co-prime $dG $ and $dP_N/y$.

Hence, we proved that the algebra of forms (\ref{algforms})
exists and is associative for the Seiberg-Witten
theory with the data (\ref{suncu})
and thus $C_{ij}^k(dG )$ satisfy the
associativity condition
\be
{\sf C}_i(dG ){\sf C}_j(dG ) = {\sf C}_j(dG ){\sf C}_i(dG )
\ee
This is always the case for the Seiberg-Witten theories associated with
hyperelliptic curves, algebra of forms (\ref{algforms}) there is
isomorphic to some
polynomial ring and isomorphism is basically given by relation
(\ref{holdif}). This is much less trivial in the case of non-hyperelliptic
curves. Several examples of the closed associative algebras of forms are
known in these cases as well, see for example \cite{HKM}. However, as we see
below in sect.~\ref{ss:consisw}, there is no special reason for the
existence of
closed and associative algebra of forms in generic situation. This fact was
first established in \cite{MMMlong} where it was shown explicitly that
algebra of holomorphic differentials for the Seiberg-Witten theory
associated with the
Calogero-Moser system is {\em not} associative.

As it becomes clear from general consideration of sect.~\ref{ss:wdvvgen}
(and it will be demonstrated again for Seiberg-Witten theory in
sect.~\ref{ss:consisw}) algebra of forms is not the simplest way to study the
problem. One {\em can} introduce algebra of forms following \cite{MMMforms},
but it is much easier to consider always the ring of functions, which is
{\em always} associative. Then, the problem is again reduced to a system of
linear equations, {\em if} there exists a residue formula, and the existence
of residue formulas in Seiberg-Witten theory follows from the
arguments of sect.~\ref{ss:ressw}.

\subsection{Consistency for the Seiberg-Witten prepotentials
\label{ss:consisw}}

Let us now discuss the residue formulas for the Seiberg-Witten theory within
the general context of sect.~\ref{ss:finite}. It becomes very important,
that derived originally in the form of (\ref{v}) sum over $2g+2$ branch
points of hyperelliptic Riemann surface (\ref{c2branch}), it may be
nevertheless rewritten in the form (\ref{v1}) or, moreover
\be
\label{ressw}
{\cal F}_{ijk} =
\oint_{{dw\over w}=0} {d\omega_id\omega_jd\omega_k\over d\lambda{dw\over w}}
= \sum_\alpha \res_{\lambda_\alpha}{d\omega_id\omega_jd\omega_k\over
d\lambda {P_N'(\lambda)d\lambda\over y}} = \\ =
\sum_\alpha
\res_{\lambda_\alpha}{\phi_i(\lambda)\phi_j(\lambda)\phi_k(\lambda)
\over y^2P_N'(\lambda)} =
\sum_\alpha \res_{\lambda_\alpha}{\phi_i(\lambda)\phi_j(\lambda)
\phi_k(\lambda)\over (P_N(\lambda)^2-4)P_N'(\lambda)}
\ee
where the role of hamiltonians $\{ dH_i\}$ from (\ref{res})
is played by the set of canonical holomorphic differentials
(\ref{holdif}) on Riemann surface (\ref{suncu}), (\ref{c2}).
Note, that passing from (\ref{v}) to (\ref{ressw}) we first replaced the sum
over $2g+2$ points $\lambda_I$ in (\ref{v}) by the sum over $2g$ points (the
zeroes of $dw/w$) in (\ref{v1}), and then, restricted it further in
(\ref{ressw}) over the sum of $g=N-1$ points $\lambda_\alpha$,
which are the zeroes of ${dw\over w} = {dP_N(\lambda)\over y}$,
(or better the projections of these zeroes to the $\lambda$-plane),
{\it i.e.} solutions to $P_N'(\lambda_\alpha)=0$, using the symmetry between
two sheets of hyperelliptic curve (\ref{c2}).
The set of the polynomials $\{\phi_i(\lambda)\}$ is supposed to be
nondegenerate in
the points $\{\lambda_\alpha\}$.

We know, that despite the difference between formulas (\ref{res}) and
(\ref{v}) the WDVV equations (\ref{WDVV}) do hold for the Seiberg-Witten
theory \cite{MMM}. The reason becomes clear after we passed from (\ref{v})
to (\ref{ressw}) -- as in sect.~\ref{ss:finite},
matching condition between the number of holomorphic
differentials $d\omega_i$ and the number of zeroes of
$P_N'(\lambda)$ holds {\em exactly} since both numbers are
equal to the genus of Riemann surface (\ref{suncu}),
(\ref{c2}) which is $g=N-1$.
And this is {\em all} (together with (\ref{det})) we need for derivation
of (\ref{WDVV}) from (\ref{ressw}), hence the proof for the pure \N2 SUSY
Yang-Mills theories, and moreover, for any Seiberg-Witten theories with
hyperelliptic curves, almost literally repeats the generic proof of
sect.~\ref{ss:finite}.

Now, why this may cause difficulties in generic (non hyperelliptic) situation
in Seiberg-Witten theory \cite{MMMlong}? The reason is that, by derivation,
formula (\ref{ressw}) is {\em accidental}, since what was really
derived in sect.~\ref{ss:ressw} is formula (\ref{v}) or
\be
\label{ressw1}
{\cal F}_{ijk} = -
\oint_{d\lambda=0} {d\omega_id\omega_jd\omega_k\over d\lambda{dw\over
w}} =
\oint_{{dw\over w}=0} {d\omega_id\omega_jd\omega_k\over d\lambda{dw\over w}}
\ee
where the second integral (used in (\ref{ressw}) to be rewritten in the
form of (\ref{res})) is a consequence of the first due to holomorphic
properties of the integrand. Remember, that it is the first equality in
(\ref{ressw1}), or formula (\ref{v}) which was originally
obtained by differentiating the period matrix $T_{ij}=\F_{ij}$ of the
Riemann surface (\ref{suncu}), (\ref{c2}) and using relations (\ref{derram})
for the
values of canonical holomorphic differentials (\ref{holdif}) at the branch
points of hyperelliptic curve (\ref{c2branch}), defined by
$d\lambda=0$ (and {\em
not} by ${dw\over w}=0$). However, the number of hyperelliptic branch
points, as follows from (\ref{c2branch}), equals $2N$ and
the matching condition naively would {\em fail}! What saves the situation
is that, using that integrand in (\ref{ressw1}) is holomorphic, one can
rewrite the same contour integral around the zeroes
${dw\over w} = {dP_N(\lambda)\over y} =0$ (\ref{v1}), which
is still not enough, since the number of zeroes of ${dw\over w}$ is
$2g=2(N-1)$, but due to hyperelliptic ${\bf Z}_2$-symmetry of exchanging
$\lambda$-sheets one can finally bring the residue formula to the form of
(\ref{ressw}), {\it i.e.} as a sum over $g=N-1$ zeroes of
the polynomial $P_N'(\lambda)$ (each of them corresponds in fact to the
pair of points on curve (\ref{suncu}), (\ref{c2})). It means that the matching
condition for the Seiberg-Witten Toda chain case finally holds, but not
automatically!

Now, it becomes clear from this derivation
that for other prepotentials on Riemann
surfaces the matching conditions may fail.
For example, in the case of elliptic Calogero-Moser or
broken ${\cal N}=4$ Seiberg-Witten theory the generating
differential, instead of $\lambda {dw\over w}$ (\ref{suncu})
is $\lambda dz$, where $dz$ is canonical holomorphic differential on base
torus and function $\lambda$ satisfies the Lax equation
\be
\label{CaMo}
\det_{N\times N}(\lambda - L(z)) = 0
\ee
with the introduced in \cite{KriCal} Lax operator
\be
\label{CMLax}
L(z) = \left(\begin{array}{cccc}
 p_1 & \Phi (q_1-q_2|z) & \ldots &
\Phi (q_1 - q_{N}|z)\\
\Phi (q_2-q_1|z) & p_2 & \ldots &
\Phi (q_2-q_{N}|z)\\
 & & \ldots  & \\
\Phi (q_{N}-q_1|z) &
\Phi (q_{N}-q_2|z)& \ldots &p_{N}
\end{array} \right)
\ee
where the matrix elements
$\Phi (q|z) = m\frac{\sigma(q+z)}{\sigma(q)\sigma(z)}
e^{\zeta(q)z}$ are expressed in terms of Jacobi sigma-functions
on base elliptic curve. The number of
zeroes of $d\lambda$ and $dz$ can be easily calculated from the Riemann-Roch
theorem, saying, in particular, that for any meromorphic differential
\be
\# ({\rm zeroes}) - \#({\rm poles}) = 2g-2 = 2N-2
\ee
since the genus of the curve
(\ref{CaMo}) is $g=N$. The differential $dz$ is holomorphic
(it is holomorphic on base torus and does not acquire poles on the cover),
so one gets
\be
\label{zedz}
\# ({\rm zeroes}\ dz) = 2N-2
\ee
quite similar to its analog ${dw\over w}$ in Toda case (\ref{suncu}),
(\ref{c2}). However, we
do not have anymore the hyperelliptic symmetry, which allows to "reduce factor
$2$" and, say, rewrite
(\ref{ressw}) as a sum over only $(N-1)$ points ({\em half} of
$(2N-2)$). As for the second differential $d\lambda$, since it follows from
(\ref{CaMo}) and the properties of the elliptic Calogero-Moser Lax
operator (\ref{CMLax}) that
\be
d\lambda \sim {dz\over z^2}
\ee
it has $N$ second-order poles, hence
\be
\# ({\rm zeroes}\ d\lambda) - \#({\rm poles}\ d\lambda) =
\# ({\rm zeroes}\ d\lambda) - 2N = 2N-2
\ee
or
\be
\# ({\rm zeroes}\ d\lambda) = 4N-2
\ee
{\it i.e.} it has even more zeroes than (\ref{zedz}). It means that restricting
(\ref{res}) to the case of holomorphic differentials only
\be
\label{ressw2}
{\cal F}_{ijk} = -
\oint_{d\lambda=0} {d\omega_id\omega_jd\omega_k\over d\lambda dz} =
\oint_{dz=0} {d\omega_id\omega_jd\omega_k\over d\lambda dz}
\ee
one would get
$\#(\alpha)>\#(i)$ and in order to
close the algebra, corresponding to (\ref{ressw2})
one needs to add to the set of $N$ holomorphic
differentials at least $(N-2)$ extra ``hamiltonians''. Naively there are
two direct options to do that -- to add either meromorphic differentials
or non single-valued holomorphic differentials in spirit
of \cite{KriW}. One needs then, however, to check the (extended) residue
formula (\ref{ressw2}) with added new meromorphic or non single-valued
differentials.
From the point of view of SUSY quantum theory the main problem is the
physical sense of corresponding extra time variables which should play the
role of "hidden" moduli parameters in corresponding Seiberg-Witten theory.

\section{Duality and WDVV equations}
\label{ss:wdvvdual}

\subsection{Special K\"ahler geometry and S-duality}

Duality transformations play an important role in modern theoretical
physics. In Seiberg-Witten theory \cite{SW} electric-magnetic duality
is a basic
ingredient in obtaining the exact form of the low-energy effective
action. Hence, duality is a crucial tool in studying
non-perturbative physics. Any truly non-perturbative
result should be consistent with electric-magnetic duality.

Based on electric-magnetic duality, the Seiberg-Witten theory enables the
determination of the holomorphic function $\F({\bf a})$ (\ref{prepsw})
in terms of
which the low-energy effective action is encoded.
The function $\F$ plays the role of a prepotential for the
corresponding special K\"ahler geometry. The construction
involves an auxiliary complex curve, whose moduli space of complex
structures is  identified with the special K\"ahler space\footnote{
  Strictly speaking `special geometry' refers to the K\"ahler
  geometry associated with locally ${\cal N}=2$ supersymmetric Yang-Mills
  theories coupled to Poincar\'e supergravity. In the rigid supersymmetry
  context one sometimes uses the term `rigid special geometry'. Here we
  do not make this distinction.} 
with $\bf a$ playing the role of local coordinates. This
construction can be cast in terms of an integrable system \cite{GKMMM},
identifying $\F({\bf a})$ with (the logarithm of) a tau-function of
the so-called quasiclassical or universal Whitham hierarchy \cite{KriW}, which
satisfies a set of nontrivial differential equations (see, for
example, \cite{Mbook,Mtmf} and references therein for the details of this
correspondence).

In Seiberg-Witten theory the second derivatives of ${\cal F}({\bf a})$
(\ref{prepsw})
are identified with the period matrix (\ref{pemat}) of the corresponding
auxiliary complex curve, for example (\ref{suncu}) or (\ref{CaMo}).
It's imaginary part is equal to the K\"ahler metric, as follows
directly from the K\"ahler potential,
\be
K({\bf a},\bar{\bf a}) = \Im\; \left(\sum_i\,{\bar a}_i\,{\d\F\over\d
a_i}\right)
\label{K-pot}
\ee
The derivatives of the period matrix, $\d T_{ij}/\d a_k = \F_{ijk}$,
are the (totally symmetric) holomorphic tensors (\ref{matrF})
that appear in the
WDVV equation (\ref{WDVV}).

According to electric-magnetic duality two dual
holomorphic functions, describe the same geometry and
thus belong to the same equivalence class. As it was already stressed, this
duality is therefore at the basis of the Seiberg-Witten
theory. Consequently it follows that physically relevant results, when
expressible directly in terms of the function $\F$,
should hold for all representatives of the equivalence
class. Specifically, all functions from the equivalence class of
$\F$ should both satisfy the
corresponding relations. Therefore
it follows that the associativity equations should hold for all
representatives of a given equivalence class.

The issue of the metric that appears in the associativity equations
remains a confusing one. I have already stressed that the special
K\"ahler metric has nothing in common with the ``metric'' in the
context of the two-dimensional topological theory that underlies the original
WDVV equations. The latter is related to the third derivatives of function
$\F$ (\ref{metric}) and for two-dimensional topological theory
can be chosen constant.
Note that the extra condition of
the constancy is {\em not} preserved under duality, so that duality
seems to take us out of the class of topological solutions in the sense
of \cite{Dub}. The first metric, on the other hand, is related to the
second derivative and it is non-holomorphic and transforms
non-holomorphically under duality (cf. with the formula (\ref{G-dual})
below). In
contradistinction with the above, the
metric in the context of the associativity equations (\ref{WDVV}) is
clearly non-constant and holomorphic and, as was already mentioned,
this is a very important issue.

The covariance of WDVV equations w.r.t. duality transformations goes
beyond Seiberg-Witten theory because it
applies to all
cases where the equations (\ref{WDVV}) are valid, irrespective of whether one
can identify proper arguments for the relevance of electric-magnetic
duality for the cases at hand. All Seiberg-Witten solutions are related
to integrable systems,
where the function $\F$ is the logarithm of the tau-function of the
universal Whitham hierarchy (restricted to a finite set of variables)
\cite{GKMMM}, see also \cite{Mbook} and references therein.
However, not all tau-functions correspond to Seiberg-Witten solutions,
and some of those nevertheless satisfy the WDVV
equations (\ref{WDVV}). Hence, by applying duality transformations we
obtain other tau-functions satisfying the WDVV equations, without
having an a priori understanding as to why the duality constitutes an
equivalence relation for these tau-functions.

Yet another issue concerns the relation of WDVV equations in
Seiberg-Witten theory with the geometry of moduli spaces of Riemann
surfaces and integrable systems. Certainly dual period matrices are
not distinguishable from the point of view of the geometry of complex
curves. They are equivalent and the corresponding equivalence of the
associativity equations is a consequence of this fact. On the other
hand, it is well-known that when
two different functions $F({\bf a})$ and ${\tilde F}(\tilde{\bf
a})$ satisfy
\be
{\d^2 F\over \d a_i\,\d a_j} = {\d^2 {\tilde F}\over \d {\tilde a}_i\,
\d {\tilde a}_j}
\label{DubMiMo}
\ee
and $F({\bf a})$ is a solution to WDVV equations, then ${\tilde
F}(\tilde{\bf a})$ is trivially a solution to the same equations. In
\cite{dWM} this equivalence was extended to the case of functions whose
second derivatives ({\it i.e.} their period matrix) are related by duality
transformations. Observe that, while representing the same geometry
and belonging to the same equivalence class, the two functions which
solve the WDVV equations are in general completely different as
functions depending on their respective arguments.

\subsection{Electro-magnetic duality and WDVV equations}

Following \cite{dWM}, let is now consider the electric-magnetic
duality transformation
\be
\label{ad}
a_i\to a^D_i = {\d\F\over\d a_i}\,, \qquad a_i^D \to - a_i = {\d\F^D({\bf
a}^D)\over \d a^D_i}
\ee
with the dual function $\F^D({\bf a}^D)$. As is well-known, this
transformation is effected by a Legendre transform,
\be
\F^D({\bf a}^D) = \F({\bf a}) - \sum_i \, a_i\,a^D_i
\label{Legendre}
\ee
Obviously we have
\be
{\d a^D_i\over \d a_j}= {\d^2\F\over\d a_i\,\d a_j} = T_{ij}\,, \qquad
{\d a_i\over \d a^D_j}=  - {\d^2\F^D\over\d a^D_i\,\d a^D_j}= -
T^D_{ij}
\ee
so that the dual period matrix $T^D_{ij}$ equals minus the inverse of
the original period matrix (\ref{pemat}), {\it i.e.}
\be
\label{ttd}
\sum_j \,T^D_{ij}\, T_{jk} = - \delta_{ik}
\ee
Consider now
\be
\label{fd3}
\|\f^D_i\|_{jk} \equiv \F^D_{ijk}= - {\d T^D_{ij}\over\d a^D_k} =
{\d^3\F^D\over\d a^D_i\,\d a^D_j\,\d a^D_k}
\ee
It directly follows that
\be
{\d T^D_{ij}\over\d T_{kl}} =  T^D_{ik}\ T^D_{lj}
\ee
and
\be
{\d T^D_{ij} \over \d a^D_k} = \sum_{l,m,n}\, T^D_{il} \,{\d T_{mn}\over \d
a_l}\,T^D_{nj} \;  {\d a_l \over \d a^D_k}
\ee
Consequently $\F_{ijk}$ transforms just as the third derivative of a
function under the corresponding (linear) reparametrization,
\be
{\d^3\F^D\over\d a^D_i\,\d a^D_j\,\d a^D_k}= \sum_{l,m,n} \,
{\d^3\F\over\d a_l\,\d a_m\,\d a_n} \;  {\d a_i \over \d a^D_l} \,{\d
a_j \over \d a^D_m}\, {\d a_k \over \d a^D_n}
\label{3df-id}
\ee
or, in matrix form,
\be
\|\f^D_i\|  = \sum_j \, {\d a_i \over \d a^D_j} \; \|T^D\cdot \f_j\cdot
T^D \|
\ee
From this result it is obvious that the equations (\ref{WDVV}) are
valid for the dual function $\F^D({\bf a}^D)$, because
\be
\f^D_i\cdot(\f^D_j)^{-1}\cdot\f^D_k - (i\leftrightarrow k)  = \sum_{l,m,n}
{\d a_i \over \d a^D_m}\,{\d a_k \over \d a^D_n}\,{\d a^D_j \over \d a_l}
\,\left(\f_m\cdot\f_l^{-1}\cdot\f_n - (m\leftrightarrow n) \right) =0
\label{WDVV-eq}
\ee
where in the r.h.s. we made use of (\ref{WDVV}) for all
$l,m,n$.

The same logic can be applied to generic electric-magnetic
duality transformations forming an arithmetic subgroup of $Sp(2r,{\bf
R})$, which generalize the special duality transformation given in formula
(\ref{ad}). Here $r$ is the rank of the gauge group. At the
perturbative level these transformations are continuous.
The covariance properties established in \cite{dWM} and discussed below
do not depend on this feature.

In the dual basis (denoted by the superscript $S$), we have new
variables ${\bf a}^S$ and a new function $\F^S({\bf a}^S)$, defined
by,
\be
{\bf a}^S = U\cdot {\bf a} + Z\cdot \left({\d\F\over\d{\bf
a}}\right)\,, \nonumber\\
\left({\d\F^S\over\d{\bf a}^S}\right)=  V\cdot
\left({\d\F\over\d{\bf a}}\right)+ W\cdot {\bf a}
\ee
where the $r\times r$ matrices $U$, $V$, $W$, $Z$ combine into an
$Sp(2r,{\bf Z})$ matrix
\be
{\cal O} = \left(
\begin{array}{cc}
  U & Z  \\
  W & V
\end{array}\right)
\ee
by virtue of relations
\be
\label{uvwz}
U^t\cdot V-W^t\cdot Z = V\cdot U^t-Z\cdot W^t = 1
\\
U^t\cdot W = W^t\cdot U\,,\qquad Z^t\cdot V = V^t\cdot Z
\ee
which mean that matrix ${\cal O}$
has unit determinant and obeys
\be
\label{oomega}
{\cal O}^{-1} = \Omega{\cal O}^t\Omega^{-1}
\ee
where $\Omega$ is $2r\times 2r$ matrix
\be
\Omega =
\left(\begin{array}{cc}
  0 & 1  \\
  -1& 0
\end{array}\right)
\ee
The result analogous to (\ref{Legendre}) reads,
\be
\F^S ({\bf a}^S) =  \F({\bf a})  + {1\over2} {\bf
a}^t\cdot U^t \cdot W\cdot {\bf a} + {\bf a}^t\cdot W^t \cdot Z \cdot
\left({\d\F\over\d{\bf a}}\right)  + {1\over2}
\left({\d\F\over\d{\bf a}}\right)^t\cdot Z^t \cdot V\cdot
\left({\d\F\over\d{\bf a}}\right)
\label{fsdu}
\ee
Observe that this represents only a (partial) Legendre transform when
$U^t\cdot W = Z^t\cdot V=0$.

From these results one proves that the period matrix, again defined
by (\ref{pemat}), and its dual counterpart in general situation
\be
{\d^2\F^S\over\d a^S_i\,\d a^S_j} = T^S_{ij}
\ee
are related by
\be
\label{gendual}
T^S = (V\cdot T + W)\cdot S^{-1}(T)
\ee
where matrix $S(T)$ is defined by
\be
\label{S}
S_{ij}(T) = {\d a_i^S\over \d a_j} = \|U + Z\cdot T\|_{ij}
\ee
The special K\"ahler metric associated with
the K\"ahler potential (\ref{K-pot}),
\be
G_{\bar \imath j} = {\d^2 K({\bf a},\bar{\bf a})\over \d \bar
a_{\bar\imath} \,\d a_j} = \Im\; T_{ij}
\label{Kahme}
\ee
transforms as
\be
G^S= [S^\dagger]^{-1}(\bar T) \cdot G  \cdot S^{-1}(T)
\label{G-dual}
\ee
Now let us demonstrate that
the {\em third} derivatives of $\F$ and $\F^S$ remain related just as in
(\ref{3df-id}), {\it i.e.},
\be
\label{srel}
\F^S_{ijk} =\sum_{l,m,n} \, \F_{lmn}
\,(S^{-1})_{li}\,(S^{-1})_{mj}\,(S^{-1})_{nk}
\ee
or
\be
{\d^3\F^S\over\d a^S_i\,\d a^S_j\,\d a^S_k}= \sum_{l,m,n}\,
{\d^3\F\over\d a_l\,\d a_m\,\d a_n} \;
{\d a_i \over \d a^S_l} \,{\d a_j \over \d a^S_m}\,
{\d a_k \over \d a^S_n}
\label{3sf-id}
\ee
First one shows that
\be
\label{dts}
\delta T^S = \left(V - (V\cdot T+W)\cdot S^{-1}\cdot Z\right)\cdot
\delta T\cdot S^{-1} = (S^t)^{-1}\cdot\delta T\cdot S^{-1}
\ee
This result (\ref{dts}) follows directly from the equations
(\ref{gendual}), (\ref{S}) and (\ref{uvwz}). Likewise one shows
that $S^{-1}(T)\cdot Z$ is a symmetric matrix.

Replacing the variation in (\ref{dts}) by a derivative with
respect to ${\bf a}^S$ and using $\F_{ijk} = \d T_{ij}/ \d a_k$ and
(\ref{S}), one readily proves the
validity of (\ref{srel}). Along the same line as above,
this then leads to the conclusion that
the WDVV equations (\ref{WDVV}) remain covariant under {\it general}
duality transformations (\ref{gendual}), so that the function
$\F^S({\bf a}^S)$ satisfies
\be
\label{WDVVS}
\f^S_i\cdot (\f^S_j)^{-1}\cdot\f^S_k = \f^S_k\cdot
(\f^S_j)^{-1}\cdot\f^S_i
\ee
provided the WDVV equations were valid for the original function
$\F$. Upon setting $U=V=0$ and $Z=-W=1$, the reader can also verify that
formulas (\ref{WDVV-eq}) are reproduced as a particular case.

The formula (\ref{dts}) is a simple consequence of duality transformation
(\ref{gendual}) and its transponed version with $T^t=T$
and $(T^S)^t = T^S$. From (\ref{dts}) and (\ref{S}) it follows that
\be
\label{deltr}
\delta T^S = \left(V - (V\cdot T+W)\cdot S^{-1}\cdot Z\right)\cdot
\delta T\cdot S^{-1} = \\ =
(S^t)^{-1}\cdot\delta T\cdot \left(V^t - Z^t\cdot
(S^t)^{-1}\cdot (T\cdot V^t + W^t) \right)
\ee
and one can easily prove equivalence of (\ref{deltr}) and (\ref{dts}) using
relations (\ref{uvwz}).
Duality transformations
with $U^t\cdot W = Z^t\cdot V=0$, where (\ref{fsdu}) takes the form of
a (partial) Legendre transform,  may be of particular importance in
the context of the Whitham hierarchies.


\section{Associativity equations in dispersionless
               integrable hierarchies}

\subsection{WDVV equations from Hirota relations}

One may also consider the WDVV equations (\ref{WDVV})
when the number of variables is infinite, as we see immediately in this
case the form (\ref{WDVV2}) is mostly adequate. Moreover, following
\cite{confmaps,BMRWZ},
it is convenient to use generating functions for the
derivatives of ${\cal F}$.
Introducing operator
\be
\label{eq:3}
D(z) = \sum_{k=1}^\infty  \frac{z^{-k}}k \frac \d{\d t_k}
\ee
one may define the generating functions for the second (\ref{eq:2})
\be
\label{eq:4}
D_1D_2{\cal F} \equiv D(z_1)D(z_2){\cal F} = \sum_{k, m=1}^\infty
\frac{z_1^{-k}}{k} \frac{z_2^{-m}}{m} \F_{km}
\ee
and third (\ref{matrF})
\be
\label{eq:61}
D_1D_2D_3{\cal F}
\equiv D(z_1)D(z_2)D(z_3){\cal F} = \sum_{k, m, n=1}^\infty
\frac{z_1^{-k}}k \frac{z_2^{-m}}m \frac {z_3^{-n}}n \F_{kmn}
\ee
derivatives. Following \cite{BMRWZ} we also introduce
generating functions for the structure constants (\ref{ceta}), (\ref{cff})
\be
      \label{eq:14}
       C^l (z_1 , z_2 )= \sum_{i, j=1}^\infty C^l_{ij}
\frac{z_1^{-i}}i \frac{z_2^{-j}}j
\ee
and for the $X_{ijkn}$~(\ref{x})
\be
   \label{eq:16a}
   X(z_1,z_2,z_3,z_4) \equiv
\sum_{i,j,k,n=1}^{\infty}
     \frac{z_1^{-i}}{i} \frac{z_2^{-j}}{j} \frac {z_3^{-k}}{k}
\frac{z_{4}^{-n}}{n}
X_{ijkn}.
\ee
Our starting point in this section is the bilinear identity for the
tau-function which we refer to as Hirota equation.  Let
${\cal F}$
be the dispersionless limit of logarithm of the KP tau-function ${\cal
  F}\equiv \log\tau$~\cite{KriW,TakTak}, in this limit the
Hirota equations encode the set of relations for the second order
derivatives ${\cal F}_{ij}$. In generating form they can be written as
\cite{TakTak,KodamaCarrol}:
\be
      \label{eq:7}
      (z_1 - z_2) \left (1- e^{D_1D_2 {\cal F}}\right ) =
\left (\strut D_ 1 - D_
        2\right )  \d_{t_1}{\cal F}
\ee
where we use the operator~(\ref{eq:3}). The symmetric
version of this equation is
\be
      \label{eq:8}
      (z_1-z_2)e^{D_1D_2 {\cal F}} +
      (z_2-z_3)e^{D_2D_3 {\cal F}} +
      (z_3-z_1)e^{D_3D_1 {\cal F}} = 0
\ee
Note that one can obtain~(\ref{eq:7}) from~(\ref{eq:8}) in the limit
$z_3\rightarrow \infty$.
These equations should be understood as an infinite set
of algebraic relations for ${\cal F}_{ij}$ obtained
by expanding both sides of equalities as power series in
(inverse degrees of) $z_i$ and comparing the corresponding coefficients.
These relations can be
resolved with respect to ${\cal F}_{ij}$ with $i,j \geq 2$.
Indeed, writing~(\ref{eq:7}) as
\be
\label{17}
D(z_1)D(z_2){\cal F}=\log \frac{w(z_1)-w(z_2)}{z_1-z_2}
\ee
where
\be
\label{eq:10}
w\equiv w(z)=z-\sum_{k=1}^\infty\frac{z^{-k}}{k}{\cal  F}_{1k}
\ee
one may conclude that
\be
\label{hirp}
{\cal F}_{ij}=P_{ij}({\cal F}_{11},\,
{\cal F}_{12},\, {\cal F}_{13}, \, \ldots \,)
\ee
with $P_{ij}$ being certain polynomials.

Second order derivatives of the tau-function allow
to define a set of commuting
flows with generators $H_k$ determined from the series
\be
\label{hams}
     D(z_1)D(z_2){\cal F}=-\log \left (1-\frac{z_2}{z_1}\right )-
\sum_{k=1}^{\infty}
     \frac{z_1^{-k}}{k}H_k(z_2)
\ee
Acting by $D(z_3)$ on both sides and interchanging $z_1$
and $z_3$ one finds that
\be
\label{commute}
\frac{\d H_i(z)}{\d t_j}=\frac{\d H_j(z)}{\d t_i}
\ee
Note that $H_1(z)=w(z)$~(\ref{eq:10}). The
relations~(\ref{commute}) can be viewed as a hierarchy
of evolution equations for the $w(z)$
\be
\label{weq}
\frac{\d w(z)}{\d t_k}=
\frac{\d H_k (z)}{\d t_1}
\ee
Equations~(\ref{weq}), being rewritten as
evolution equations for the function $z(w)$, have the form of
dispersionless Lax equations
\be
\label{Lax}
\frac{\d z(w)}{\d{t_k}}=
\{H_k(w),\, z(w)\}_{\rm KP}
\ee
where the Poisson brackets are defined as
\be
\{f,g\}_{\rm KP}=
\frac{d f}{d w}\frac{\d g}{\d t_1}
-\frac{\d f}{\d t_1}\frac{d g}{d w}
\ee
and the derivatives in $t_i$ are taken at fixed $w$.
Moreover, as it follows from~(\ref{17}),
the $H_k$ turn out to be polynomials in $w$. On the other hand,
(\ref{hams}) fixes $H_k$ to be of the form
$H_k =z^k -D(z)\d_{t_k}{\cal F}$, {\it i.e.}
$H_k =z^k + O(z^{-1})$.
Therefore
\be
\label{f}
H_k=(z^k(w))_{\geq 0}
\ee
where the symbol $(f(w))_{\geq 0}$ means the non-negative part
of the Laurent series in $w$, this is the dKP hierarchy (see
e.g.~\cite{TakTak}).
Given a Lax function $z(w)=w+O(z^{-1})$ and $H_k$ obtained
from it by means of~(\ref{f}), one can reconstruct the
second order derivatives ${\cal F}_{jk}$ via the formula
\be
\label{res2ord}
{\cal F}_{jk}=\frac{1}{j+k}\,\mbox{res}_{\infty}
\left ( \frac{dH_j dH_k}{d \log z}\right )
\ee
which is an analog of (\ref{2derLG}) and even, in some sense of
(\ref{prepsw}), (\ref{pemat}).
Note that eqs.~(\ref{hams}) -- (\ref{Lax}) hold for
any function ${\cal F}$, but this is not an integrable
hierarchy yet, in spite of the fact that there are infinitely many
commuting flows~(\ref{commute}). The crucial relation, which really makes an
integrable hierarchy out of this, is~(\ref{f})
\footnote{In the case of quasiclassical hierarchy of "general position"
one should impose onto hamiltonians -- analogs of (\ref{f}) (more exactly, onto
their differentials) certain analytic properties on some Riemann surface.}.
The Hirota equation gives a relation between the generating function of
the flows and (globally defined) function $w$ and allows one to
determine $H_k$ as functions $H_k(w)$ with certain analytic properties,
in the
dKP-case they are polynomials. From this point of view, it is the Hirota
identity that encodes integrability of the system.

Plugging~(\ref{17}) in the r.h.s. of~(\ref{hams}) and differentiating
w.r.t.
$w=w(z_2)$, one arrives to the relation
\be
\label{gener}
\frac{1}{w(z_1)-w}=\sum_{k\geq 1} \frac{z_{1}^{-k}}{k}\
\frac{d H_k (w)}{d w}
\ee
which is used below to get an explicit realization of associative
algebra.

Let us stress that all basic relations of the dKP (and dToda) hierarchy
contain second order derivatives of ${\cal F}$ only.
An elementary manipulation with the Hirota equations~(\ref{eq:7}) allows
to
bring them to the form~(\ref{cf}), which is the defining relation for the
structure constants $C_{ij}^{l}$.  Applying $D(z_3)$ to both sides
of~(\ref{eq:7}) gives
\be
      \label{eq:11}
D_1 D_2 D_3 {\cal F} = -\, \frac{1}{w_1 - w_2}(D_1 -
        D_2)D_3\d_{t_1}{\cal F} = -\ {e^{-D_1D_2\F}\over z_1-z_2}(D_1 -
        D_2)D_3\d_{t_1}{\cal F}
\ee
where $w_i \equiv w(z_i)$. Notice immediately
that eq.~(\ref{eq:11}), being written in components
using~(\ref{eq:3}), (\ref{eq:61}) and (\ref{eq:14}),
is equivalent to
the infinite-dimensional version of~(\ref{cf}):
\be
      \label{eq:15}
      \F_{ijk} = \sum_{l=1}^\infty C^l_{ij} \F_{lk1}
\ee
where the structure constants are defined by the
generating function (\ref{eq:14})
\be
      \label{eq:142}
      C^l(z_1 , z_2 )=
      -\, \frac{z_1^{-l} - z_2^{-l}}{l(w_1 - w_2)} =
      - \, \frac{z_1^{-l} - z_2^{-l}}{l(z_1 - z_2)}
\, e^{-D(z_1)D(z_2) {\cal F}}
\ee
It is easy to see from~(\ref{eq:14}) that the infinite
sum in~(\ref{eq:15}) is actually always finite: it truncates
at $l=i+j$.

Let us show that ${\cal F}$, defined by (\ref{eq:7}) and/or (\ref{eq:8}),
obeys the WDVV equations (\ref{WDVV2}), with each index running
over the infinite set of natural numbers.
In terms of generating functions this means that
$X(z_1, z_2 , z_3 , z_4 )$ given by~(\ref{eq:16a})
is totally symmetric w.r.t. permutations of $z_1 \dots z_4$.
It is enough to prove the symmetry w.r.t. the permutation
of $z_2$ and $z_3$, which is equivalent to the relation
\be
   \label{eq:18}
   z_{13}e^{D_1D_3{\cal F}}(D_1-D_2)D_3D_4{\cal F} =
z_{12}e^{D_1D_2{\cal
       F}}(D_1-D_3)D_2D_4{\cal F},
\ee
where $z_{ik}=z_i - z_k$, or
\be
\label{eq4}
z_{13}e^{D_1D_3\F }D_1D_3D_4\F  - z_{12}e^{D_1D_2\F }D_1D_2D_4\F  =
z_{13}e^{D_1D_3\F }D_2D_3D_4\F  - z_{12}e^{D_1D_2\F }D_3D_2D_4\F  =
\\ =
D_2D_3D_4\F \left(z_{13}e^{D_1D_3\F } - z_{12}e^{D_1D_2\F }\right)
\ee
Using~(\ref{eq:8}) it is straightforward to bring~(\ref{eq4})
into the form
\be
\label{eq:22}
D_4 \biggl(z_{13}e^{D_1D_3{\cal F}} - z_{12}e^{D_1D_2{\cal
     F}} - z_{23}e^{D_2D_3{\cal F}} \biggr) = 0
\ee
which is the $D_4$-derivative of~(\ref{eq:8}) and therefore
the WDVV equations (\ref{WDVV2}), (\ref{WDVV}) follow from the
Hirota equations.

One can prove in a similar way the WDVV equations for ${\cal F}$
choosing $D(z_a){\cal F}_{ij}$ rather than ${\cal F}_{1ij}$
as a "metric" and restoring symmetry of (\ref{WDVV}).
Applying $D(z_3)$ to the symmetric form of the Hirota equations
(\ref{eq:8}) written for the three points $z_1$, $z_2$, $z_a$
one gets
\be
\label{sym}
D_1D_2D_3{\cal F} = \frac{1}{w_{12}}
(w_{1a}D_1 - {w_{2a}}D_2)D_3 D_a {\cal F},
\ee
where
\be
\label{wik}
w_{ik}\equiv (z_ i - z_ k)e^{D_i D_k {\cal F}}
\ee
is a sort of inverse "dressed two-point correlator".
Similarly to the previous case, which is reproduced in the
limit $z_a \to \infty$, equality (\ref{sym}) defines the structure
constants.
The WDVV equations are then equivalent to
\be
\label{eqg}
w_{1a}\left(w_{13}D_1D_3D_4{\cal F} - w_{12}D_1D_2D_4{\cal F}\right) =
\left(w_{13}w_{2a} - w_{12}w_{3a}\right)D_2D_3D_4{\cal F}
\ee
Plugging the $D_4$-derivative of~(\ref{eq:8}),
write the l.h.s. of~(\ref{eqg})  as
$w_{1a}w_{23}D_2D_3D_4{\cal F}$. It is clear then that~(\ref{eqg}) is
equivalent to the identity
\be
w_{1a}w_{23} = w_{13}w_{2a} - w_{12}w_{3a}
\ee
which is automatically satisfied by (\ref{wik}) since
$w_{ij} = w_i - w_j$.

\subsection{Finite from infinite}

To give a realization of the associative algebra with the
structure constants defined by~(\ref{eq:142}), we introduce
the polynomials
\be
\label{basis}
\phi_k (w) =\frac{d H_k (w)}{d w}\,,
\;\;\;\;\;\;k\geq 1.
\ee
Expanding both sides of the identity
\be
\frac{1}{(w-w_1)(w-w_2)}=\frac{1}{w_1 -w_2}\left(
\frac{1}{w-w_1}-\frac{1}{w-w_2}\right)
\ee
in $z_1^{-1}$, $z_2^{-1}$,
using
(\ref{gener}), and comparing the coefficients,
we obtain the algebra
\be\label{alg1}
\phi_i (w) \phi_j (w) =\sum_{k\geq 1}C_{ij}^{k}\phi_k (w)
\ee
where the structure constants are exactly those defined
by~(\ref{eq:142}). In contrast to (\ref{ringw}) this infinite-dimensional
algebra is just the ring of polynomials of {\em arbitrary} degree since no
factorization (like over $dW=0$) is implied in (\ref{alg1}).

For completeness, we show how to derive the
residue formula~(\ref{res})
for third order derivatives of ${\cal F}$ directly
from the Hirota equation. Substituting into
(\ref{eq:11}) its particular case $D_1D_2\d_{t_1}{\cal F} =
-\frac{(D_1-D_2)\d_{t_1}^2{\cal F}}{w_1 -w_2}$ (obtained in the
limit $z_3\to\infty$),
one easily expresses
$D_1 D_2 D_3 {\cal F}$ in terms of $D_i \d^{2}_{t_1}{\cal F}$
only:
\be
D_1 D_2 D_3 {\cal F}=\sum_{i =1}^{3}\mbox{res}_{w_i}
\left ( \frac{D(z(w)) \d^{2}_{t_1}{\cal F}}{(w-w_1)(w-w_2)(w-w_3)}dw
\right )
\ee
In the numerator we have: $D(z)\d^{2}_{t_1}{\cal F}=
-\d w(z)/\d t_1$ which is equal to $\d_{t_1}z(w)/z'(w)$ in terms of
the independent variable $w$ ($z'(w)\equiv dz/dw$). Expanding both sides
of the above equality in series in $z^{-1}_{1,2,3}$ and using
(\ref{gener}) one obtains:
\be\label{res1}
{\cal F}_{ijk}=\frac{1}{2\pi i}\oint_{C_{\infty}}\!
\frac{\d_{t_1}z(w)}{z'(w)}
\phi_i (w) \phi_j (w) \phi_k (w) dw
\ee
where $C_{\infty}$ is a small contour around infinity in the domain
where $w(z)$ is holomorphic and univalent and $z'(w)$
does not have zeros and singularities.
Note that "topological metric" $\eta_{jk}={\cal F}_{jk1}$
(just because, as in the Landau-Ginzburg case $\phi_1 =1$).
Therefore, the algebra~(\ref{alg1}) is in full agreement
with~(\ref{cf}):
\be\label{proof}
{\cal F}_{ijk}=
\sum_{l}C_{ij}^{l}{\cal F}_{lk1} = \sum_{l}C_{ij}^{l}\eta_{kl}
\ee
If there exists a times-independent function $\varphi (z)$ such that
$E(w)\equiv\varphi (z(w))$ is a meromorphic function of $w$
with the number of poles being unchanged under variations of all
$t_j$, then the integral is equal to the sum of
residues at zeros of $E'(w)$
\be\label{res3}
{\cal F}_{ijk}=
\sum_{E'(w_{a})=0} \mbox{res}_{w_a} \left (
\frac{\d_{t_1}E(w)}{E'(w)}
\phi_i (w) \phi_j (w) \phi_k (w) dw \right )
\ee
Indeed, $\d_{t_1}z(w)/z'(w) =\d_{t_1}E(w)/E'(w)$
and poles of $\d_{t_1}E(w)$
do not contribute to the integral since they are canceled
by those of $E'(w)$.
The existence of such function $E$ leads to a finite-dimensional
reduction of the hierarchy, in this case the WDVV algebra
becomes finite-dimensional. This can be easily seen directly
from the residue formula, but now, following \cite{BMRWZ},
we show this starting from the Hirota equations.

In terms of the Hirota equation the finite-dimensional reduction
is a set of additional to (\ref{hirp}) constraints for the
second order derivatives of the logarithm of tau-function
\be
\label{eq:28}
\F_{1M} = P_{1M} (\F_{11},\dots,\F_{1,N-1})\,,
\;\;\;\;\;\; M\geq N
\ee
where functions $P_{1M}$ do not explicitly depend on times and are
required to be consistent with the evolution equations.
Any reduction of this kind leaves us with
$N-1$ independent variables -- "primary fields"
which can be chosen to be ${\cal F}_{1j}$ with $1\leq j \leq N-1$.
All other ${\cal F}_{1M}$ with $M\geq N$ ("descendants") are
expressed through the independent ones via formulas~(\ref{eq:28}).
A particular example is a familiar $N$-KdV
reduction, for which ${\cal F}$ is
independent of $t_{N},\,t_{2N},\,t_{3N},\dots$
In this case functions $P_{1M}$ are certain
polynomials with rational coefficients such that $E(w)=z^N(w)$
is a polynomial in $w$. Other reductions,
when $z^N$ is a rational function, are also known~\cite{KriW}.
In the sequel we do not refer to explicit form of~(\ref{hirp}) and
(\ref{eq:28}) and use capital letters (like $J,M,L, \ldots$) for
the "descendants", {\it i.e.} for indices larger than $N-1$.

In the case of a reduction "metric"
${\cal F}_{lk1}$ in~(\ref{eq:15}) becomes a
degenerate matrix of rank $N-1$.
Indeed, taking the derivative of~(\ref{eq:28})
w.r.t. $t_k$, we find
that $J$-th line
of the matrix ${\cal F}_{jk1}$
is a linear combination of the first $N-1$ lines.
The finite WDVV equations are obtained by means of a
projection on the nondegenerate
$N-1$-dimensional subspace.

Let us now use define the structure constants following (\ref{cf})
and substituting (\ref{eq:28}) into (\ref{hirp}), this allows
to express ${\cal F}_{Lk1}$ through ${\cal F}_{lk1}$. Indeed since
$$
\F_{ij} = P_{ij}(\F_{1k}) = P_{ij}(\F_{11},\dots,\F_{1,N-1};
P_{1L}(\F_{11},\dots,\F_{1,N-1}))
$$
one gets
\be
{\cal F}_{ijk} = {\d P_{ij}\over\d t_k}=
\sum_{l=1}^{N-1}\left (
        {\d P_{ij}\over\d\F_{1l}} +
\sum_{L} {\d P_{ij}\over\d\F_{1L}}\,\frac{\d P_{1L}}{\d {\cal F}_{1l}}
\right )
{\cal F}_{lk1}
=
\sum_{l=1}^{N-1}\left (
        C_{ij}^{l} +
\sum_{L} C_{ij}^{L}\,\frac{\d P_{1L}}{\d {\cal F}_{1l}}
\right )
{\cal F}_{lk1}
\ee
The object in brackets in the r.h.s.
\be
      \label{eq:32}
      \tilde C_{ij}^{l}
=C_{ij}^{l} +
\sum_{L} C_{ij}^{L}\,\frac{\d P_{1L}}{\d {\cal F}_{1l}}
\ee
at $i,j<N$
defines the structure constants of a finite dimensional
algebra formed by the "primary fields".
The system~(\ref{eq:15}) becomes finite:
\be
      \label{eq:33}
      \F_{ijk} = \sum^{N-1}_{l=1}
      \tilde C_{ij}^{l}
     \F_{lk1}
\ee
where the "metric" $\F_{lk1}$ is non-degenerate
on the small space ($l,\,k=1,\dots, N-1$).
The structure constants of the algebra of "primary fields" obey the
  finite-dimensional WDVV equations, in other words,
$\tilde X_{ijkm}$ defined by
      $\tilde X_{ijkm}  \equiv \sum^{N-1}_{l=1}
      \tilde C_{ij}^{l} \F_{lkm} $
is  symmetric with respect to the permutations of
the (small!) indices $i,j,k,m$.
This follows from the fact that
$\tilde X_{ijkm} = X_{ijkm}$ for $i,j,k,m<N$.

The proof is straightforward after writing
\be
      \label{eq:40}
      X_{ijkm} = \sum_{l=1}^{N-1} C^l_{ij} \F_{lkm} + \sum_{L} C^L_{ij}
\F_{Lkm},
\ee
and substituting $C^l_{ij} =
      \tilde C_{ij}^{l}
-\sum_{L} C^L_{ij} \, \d P_{1L} /\d {\cal F}_{1l}$
into (\ref{eq:40}), since
     $X_{ijkm} = \tilde X_{ijkm}
      + \sum_{L} C^L_{ij} Y_{Lkm}$,
where
$Y_{Lkm}= {\cal F}_{Lkm}-
\sum_{l<N} \frac{\d P_{1L}}{\d {\cal F}_{1l}} \F_{lkm} $
vanishes.
To see this, one can express ${\cal F}_{lkm}$ inside the sum in terms of
${\cal F}_{ij1}$ using~(\ref{eq:33})
and interchange the order of summation.

\subsection{dToda hierarchy and conformal maps}
\label{ss:toda}

The dToda
hierarchy leads to the WDVV equations in a similar manner \cite{BMRWZ}.
The independent variables of the dToda hierarchy are
$t_0,t_{\pm 1},t_{\pm 2},\dots$ and it is also convenient to
introduce two generating functions
\be
      \label{eq:48}
       w^{\pm}(z) = z \exp \left (-\frac 12 \d_{t_0}^2{\cal F}
-\d_{t_0}D^{\pm}(z){\cal F} \right )
\ee
where $D^{\pm}( z)=\sum_{k =1}^{\infty}\frac{z^{-k}}{k}\,
\frac{\d}{\d t_{\pm k}}$. In terms of generating functions the
Hirota equations for dToda hierarchy read~\cite{TakTak,confmaps}
\be
      \label{tohi}
      w^{\pm}(z_1)-w^{\pm}(z_2) =
      (z_1 -z_2)\,e^{-\frac 12 \d_{t_0}^2{\cal F}} e^{-D^{\pm}_{1}
D^{\pm}_{2}{\cal F }}
\\
1 - \frac 1 {w^+ (z_1) w^- (z_2)}= e^{-D^+_1 D^-_2 {\cal F}}.
\ee
The Hirota equations~(\ref{tohi} define the
dToda hierarchy with commuting flows generated by
\be
\label{H}
H_{\pm j}(w)=\Bigl ((z^\pm(w^{\pm 1}))^j\Bigr )_{\pm} +\frac{1}{2}
\Bigl ((z^\pm(w^{\pm 1}))^j \Bigr )_{0},\quad j\geq 1\,,
\;\;\;\;\;
H_0(w)=\log w.
\ee
Here $z^{\pm}(w)$ is the inverse function of $w^{\pm}(z)$
and $(...)_{\pm}$, $(...)_{0}$ means strictly positive, negative,
and constant part of the Laurent series, respectively.
The Lax equations read
\be
\label{T1}
\frac{\d z^{\pm}(w^{\pm 1})}{\d t_j}=\mbox{sign}\,j \,
\{ H_j (w) ,\, z^{\pm}(w^{\pm 1})\}_{\rm Toda},
\ee
where the Poisson bracket is defined as
\be
\{f,g\}_{\rm Toda}=\frac{d f}{d\log w} \frac{\d g}{\d t_0}
- \frac{\d f}{\d t_0} \frac{d g}{d\log w}
\ee
The dToda analog of the generating function~(\ref{gener})
for derivatives of $H_j$ is
\be
\label{gener1}
\frac{w^{\pm 1}}{w^{\pm}(z_1)-w^{\pm 1}}=
\pm \sum_{k\geq 1} \frac{z_{1}^{-k}}{k}\,
\frac{d H_{\pm k}(w) }{d\log w}
\ee
"Metric" (\ref{etat}) is defined to be
$\eta_{ij}={\cal F}_{0ij}$,
where the indices take all integer values,
and the structure constants (\ref{fceta}) as
\be\label{stct}
{\cal F}_{ijk}=\sum_{l=-\infty}^{\infty}
C_{ij}^{l}{\cal F}_{lk0}
\ee
From this definition we have, in particular, that
$C_{0j}^{k}=\delta^{k}_{j}$ for all $j,k$, and, of course, nothing requires
even distinguished "metric" $\F_{0ij}$ to be constant.
To find other structure constants, apply $\d_{t_k}$
to the Hirota equations~(\ref{tohi})
\be
      \label{eq:51}
      D^{\pm}_{1} D^{\pm}_{2}
\d_{t_k} {\cal F}=
\frac{1}{w^{\pm}_{1} - w^{\pm}_{2} }\,
\left (
w^{\pm}_{2} D^{\pm}_{2} {\cal F}_{k0}
- w^{\pm}_{1} D^{\pm}_{1} {\cal F}_{k0}
\right )
\\
D^{+}_{1} D^{-}_{2}
\d_{t_k} {\cal F}=
\frac{1}{w^{+}_{1} w^{-}_{2} -1 }\,
\left (
{\cal F}_{k00} +
D^{+}_{1} {\cal F}_{k0}+
D^{-}_{2} {\cal F}_{k0}
\right ).
\ee
(here and below $w^{\pm}_i\equiv w^{\pm}(z_i)$).
In complete analogy with eq.~(\ref{eq:14})
generating functions for structure constants are read from
the r.h.s. of these equations. We conclude  from~(\ref{eq:51}) that
$C_{ij}^{l}=0$ whenever $i,j$ are both positive and
$l\leq 0$
or both negative and $l\geq 0$.
If all the indices are
positive or all negative, we have:
\be
      \label{eq:56}
       \sum_{\pm i \geq 1}
       \sum_{\pm j \geq 1}
C^l_{ij}
\frac{z_{1}^{\mp i}}{i}
\frac{z_{2}^{\mp j}}{j}
=-\,\frac{
w^{\pm}_{1}z_{1}^{\mp l}-
w^{\pm}_{2}z_{2}^{\mp l}}{\pm l (
w^{\pm}_{1} -w^{\pm}_{2})}\,,
\qquad
\pm l \geq 1.
\ee
When $i$ and $j$ have different signs one can use the second equation of
(\ref{eq:51}) to get:
\be
      \label{eq:69}
       \sum_{i \geq 1}
       \sum_{j \leq -1}
C^l_{ij}
\frac{z_{1}^{-i}}{i}
\frac{z_{2}^{j}}{j}
=\frac{z_{1}^{-l}}{l (
1- w^{+}_{1}w^{-}_{2})}\,,
\qquad  l \geq 1
\\
       \sum_{i \geq 1}
       \sum_{j \leq -1}
C^l_{ij}
\frac{z_{1}^{-i}}{i}
\frac{z_{2}^{j}}{j}
=-\, \frac{z_{2}^{l}}{l (
1- w^{+}_{1}w^{-}_{2})}\,,
\qquad l \leq -1
\\
            \sum_{i \geq 1}
       \sum_{j \leq -1}
C^{0}_{ij}
\frac{z_{1}^{-i}}{i}
\frac{z_{2}^{j}}{j}
=\frac{1}{1- w^{+}_{1}w^{-}_{2} }
\ee
With this definition of the structure constants
at hand, one can prove WDVV equations for
any solution to the dToda hierarchy
\be\label{WDVVToda}
  \sum_{l=-\infty}^{\infty}
  C_{ij}^{l}{\cal F}_{lkm}=
  \sum_{l=-\infty}^{\infty}
  C_{ik}^{l}{\cal F}_{ljm}
\ee
in the same way as for the dKP-case. Details of the
proof are can be found in \cite{BMRWZ}.

The realization of the associative algebra defined by the structure
constants~(\ref{eq:56})--(\ref{eq:69}) can be written with the help
of eq.\,(\ref{gener1}) exactly in the same way as in the dKP-case.
The  generators
\be
\label{infring}
\phi_i(w) = w{dH_i\over dw}
\ee
for all integer $i$
span the ring of Laurent polynomials of arbitrary degree. In this basis
the structure constants of the algebra are given by
(\ref{eq:69}).

The derivation of the residue formulas from
Hirota relations is also parallel to the dKP-case.
Consider for simplicity the case when all the indices
are positive. We have:
\be
D^+_1 D^+_2 D^+_3 {\cal F}=\sum_{\alpha =1}^{3}\mbox{res}_{w_{\alpha}}
\left ( \frac{D^+(z^+(w)) \d^{2}_{t_1}{\cal F}}{(w-w_1)(w-w_2)(w-w_3)}dw
\right )
\ee
By virtue of~(\ref{gener1}),
this is equivalent to
\be\label{res11}
{\cal F}_{jkm}=\frac{1}{2\pi i}\oint_{C_{\infty}}\!
\frac{\d_{t_0}z^+(w)}{  z^{+}(w)'}\,
\phi_j (w) \phi_k (w) \phi_m (w) \frac{dw}{w^2},\quad
j,k,m \geq 1
\ee
where again $z^+(w)' = dz^+ /dw$.
Similar formulas can be written for non-positive indices.

As shown in~\cite{confmaps},
a particular solution $\F$ to the dToda hierarchy
describes evolution of
conformal mapping of a domain in complex plane.
This solution is specified by the reality
conditions
$t_{-k}=\bar t_k$, $w^{-}(z)=\bar w^+(z)$\footnote{Here
and below in this section bar means
complex conjugation and for any series $f(z)=\sum f_k z^k$
we set $\bar f (z)=\sum \bar f_k z^k$.}
consistent with the hierarchy, under these reality
conditions $\F$ is a real-valued function of times.

Let
$z(w)=rw+\sum_{k\geq 0}u_k z^{-k}$ be the
univalent conformal map
from the exterior of the unit circle $|w|>1$ to the exterior
of a given analytic curve $\gamma$, the normalization being fixed
by the conditions that infinity is taken to infinity and
$r$ is real and positive.
Then let us set $z^+(w)=z(w)$, $z^-(w)=\bar z(w^{-1})$; function
$w(z)=w^+(z)$ is the inverse map.
It has been shown~\cite{confmaps}
that evolution of the map is described by the dToda hierarchy with the
generators of commuting flows given by~(\ref{H}). The reality conditions
imply $H_{-j}(w)=\bar H_{j}(w^{-1})$.  The times are harmonic moments of
the exterior domain
\be
t_k=\frac{1}{2\pi i k}\oint_\gamma z^{-k}\bar zdz
\ee
with the origin assumed to be outside the domain.
The ``initial conditions'' for the
solution are given
by the dispersionless limit of the string equation:
\be\label{string}
\{z(w),\, \bar z(w^{-1})\}_{\rm Toda}=1.
\ee

In this setting the residue formula~(\ref{res11})
can be written in a more transparent form.
Since $z(w)$ maps the exterior of the unit circle in a conformal manner,
for some region in the space of $t_k$,
neither zeros of $z'(w)$ nor poles or other singularities
of $z(w)$ are in the domain $|w|>1$.
Therefore, the function under the integral in~(\ref{res11})
is regular everywhere outside the unit circle except infinity.
So, the integration contour can be taken to be the
unit circle $|w|=1$:
\be\label{res111}
{\cal F}_{ijk}=\frac{1}{2\pi i}\oint_{|w|=1}\!
\frac{\d_{t_0}z(w)}{  z'(w)}\,
\phi_i (w) \phi_j (w) \phi_k (w) \frac{dw}{w^2},\quad
i,j,k \geq 1.
\ee
The string equation~(\ref{string}) reads
\be
\label{streqto}
\frac{\d_{t_0}z(w)}{z'(w)}=
- w^2\, \frac{\d_{t_0}\bar z(w^{-1})}{\bar z'(w^{-1})}+ \,
\frac{w}{z'(w) \bar z'(w^{-1})}
\ee
where $\bar z'(w^{-1})$ is the derivative $d\bar z/dw$ taken at the
point $w^{-1}$.
Plugging (\ref{streqto}) into~(\ref{res111})
and taking into account that
the function $\d_{t_0}\bar z (w^{-1})/\bar z'(w^{-1})$
is regular inside the unit circle, we come to
\be
\label{reskri}
{\cal F}_{ijk} =
\,\frac{1}{2\pi i}\oint_{|w|=1}
\frac{\phi_i(w) \phi_j(w) \phi_k (w)}{z'(w) \bar z'(w^{-1})}
\,{dw\over w}\
= -\,\frac{1}{2\pi i}\oint_{\gamma}
\frac{dH_i dH_j dH_k}{dz d\bar z}
\ee
A more detailed analysis shows that this formula is valid,
up to an overall sign,
for all integer indices, not only for positive ones
\footnote{This formula
was first derived by I.Krichever within a different approach.}.
In the basis~(\ref{infring}) the algebra reads
$\phi_i(w)\phi_j(w) = \sum_k C_{ij}^k \phi_k(w)$,
where the structure constants
are given by (\ref{eq:56}) -- (\ref{eq:69}), the proof is the same as for
(\ref{proof}).

Comparing with the dKP residue formula~(\ref{res3}), one may say that
the
requirement of reality together with that of conformality and
univalentness
effectively defines a reduction: in both cases these conditions ensure
that the integral in~(\ref{res11}) is saturated by singularities coming
from the denominator. From this point of view, one may regard conformal maps
as an infinite-dimensional reduction of the dToda hierarchy,
and the residue formulas (\ref{res111}), (\ref{reskri}) correspond to the
localization onto the contour $|w|=1$ rather than onto finite number
of points.

In the rest of this section let us discuss
further reduction leading to new solutions \cite{BMRWZ}
of finite-dimensional WDVV equations (\ref{WDVV}).
Consider a class of conformal maps
represented by Laurent polynomials of the form
\be
\label{zpol}
z(w)=rw
+\sum_{l=0}^{N-1}u_l w^{-l}
\ee
As proved in~\cite{confmaps}, this class of functions
represents conformal maps to domains with a finite number of non-zero
moments, namely, with
$t_{k}=\bar t_k =0$ for $k>N$.
The residue formula~(\ref{res}), (\ref{reskri})
with $i,j,k \geq 0$, applied to this case reads
\be\label{resspec}
\F_{ijk} =
\sum_{\alpha=1}^{N}
\frac{\phi_i(w_\alpha) \phi_j(w_\alpha) \phi_k
(w_\alpha)w_{\alpha}^{N-1}}{W''(w_\alpha)Q'(w_\alpha)}
\ee
where, in order to identify (\ref{resspec}) with (\ref{res}),
we have formally introduced two polynomials
of degree $N$ in $w$
\be
W'(w)\equiv w^N z'(w)=
rw^N -\sum_{k=1}^{N-1}ku_k w^{N-k-1}\,,
\;\;\;\;\;
Q'(w)\equiv \bar z'(w^{-1}) =
r -\sum_{k=1}^{N-1}k\bar u_k w^{k+1}
\ee
and, as usual, $w_\alpha$ are zeros only of $W'(w)$
(which are inside the unit circle while the zeros
of $Q(w)$ are outside).
Recall that $\phi_k(w)$ for $k\geq 0$
are polynomials in $w$ of degree $k$, and in particular
$\phi_0 (w)=1$.

Now one may introduce the {\em finite-dimensional} algebra
\be\label{finalg}
\phi_j (w)\phi_k (w) =\sum_{l=0}^{N-1} C_{jk}^{l}\phi_l(w)
\;\;\;\mbox{mod}\, W'(w)\,,
\;\;\;\;\;j,k=0,1, \ldots , N-1\,
\ee
which is, again similar to the Landau-Ginzburg case, an
$N$-dimensional associative algebra
isomorphic to the ring of all polynomials factorized over
the ideal generated by $W'(w)$.
It is easy to see from the residue formula
(\ref{resspec}) that the structure
constants obey $\F_{ijk}=\sum_{l=0}^{N-1}
C_{ij}^{l}\F_{lk0}$. To do that, one
should apply~(\ref{resspec}) to the $N$-dimensional set
of flows $t_0, t_1, \ldots , t_{N-1}$.
Therefore, we conclude that (logarithm of) the
tau-function for curves~\cite{confmaps}, being restricted
to the space where all the times $t_k$, $\bar t_k$ with
$k>N$ are zero
(and $t_{N}\neq 0$ plays
a role of a parameter), provides a solution to the
finite WDVV equations (\ref{WDVV}).
More precisely,
\be\label{fF}
  \left.
\F(t_0, t_1, \ldots , t_{N-1}) \equiv
\F(t_0; t_1, \ldots , t_{N-1}, t_N, 0,0, \ldots ;
\bar t_1, \ldots , \bar t_N , 0, 0, \ldots )
\right |_{t_N \neq 0, \,\,\,\bar t_j \;\; \mbox{fixed}}
\ee
solves the WDVV system~(\ref{WDVV}) with the matrices
$({\sf F}_{i})_{jk}=
\frac{\d^3 \F}{\d t_i \d t_j \d t_k}$, $0\leq i,j,k \leq N-1$.
(let us stress that the ``antiholomorphic'' times $\bar t_k$
and the highest non-zero time $t_N$ are kept
constant under the differentiation.)

In contrast to solutions to the finite WDVV system discussed
in sect.~\ref{ss:lg}, these solutions do not allow
to switch on the higher flows (in other words there is no
"gravitational descendants" or "large phase space") since they
do not preserve the form~(\ref{zpol}), or in different words rational
conformal maps do not form a finite-dimensional reduction of dispersionless
Toda hierarchy.
Hence, we see that dispersionless tau-functions, in particular
tau-functions of analytic curves or conformal maps, give
exactly the opposite example of matching violation, compare to the
Seiberg-Witten case of sect.~\ref{ss:consisw}.
Indeed, formula (\ref{resspec}) shows that for rational maps (\ref{zpol})
we get $\#(\alpha)<\#(i)$, {\it i.e.} the matching
condition is violated into the opposite (compare to Seiberg-Witten
examples) direction. It means that (logarithm of)
tau-function of rational conformal map satisfies the WDVV equation
(\ref{WDVV}) as a function of, in fact, {\em any} (in the case of
corresponding
non-degeneracy) {\em part} of its variable whose total number is equal to
$\#(\alpha)$ or to the number of zeroes of differential of rational map.
Unfortunately in the only explicit example of ellipse (see \cite{confmaps})
there are no nontrivial WDVV equations (\ref{WDVV}) since algebra
(\ref{finalg}) is two-dimensional and we know from the very beginning that
equations (\ref{WDVV}) in such case are empty. However, if one manages
to find if not explicit, but at least closed formulas for other
tau-functions of rational maps,
their logarithms would give examples of functions, satisfying equations
(\ref{WDVV}) as functions of {\em part} of their variables.

\section{Auxiliary linear problem
\label{ss:auli}}

\subsection{Formula for prepotential and topological descendants
\label{ss:AKV}}

Let us turn now to the properties of "flat sections" of the connection
(\ref{dubrovin}) or
consider the auxiliary linear problem written as
\be
\label{aulixi}
\left({\d^2\over \d t_i\d t_j} - \zeta\sum_k C_{ij}^k
{\d\over \d t_k}\right)\Psi({\bf t};\zeta) = 0
\ee
Upon decomposition
\be
\label{expxi}
\Psi({\bf t};\zeta) = \sum_{n=0}^\infty \zeta^n\Psi_n({\bf t})
\ee
equations (\ref{aulixi}) turn into a set of recursion relations
\be
\label{recur}
{\d^2\over \d t_i\d t_j}\Psi_n({\bf t}) = \sum_k C_{ij}^k{\d\over \d t_k}
\Psi_{n-1}({\bf t})
\ee
A very simple observation, first made in \cite{AKV},
is that if one chooses $\Psi({\bf t};\zeta)$ to be a
{\em vector}-function
\be
\label{expxi1}
\Psi^m({\bf t};\zeta) = \sum_{n=0}^\infty \zeta^n\Psi^m_n({\bf t})
\ee
and fixes for $n=0$
\be
\label{n0}
\Psi^m_0({\bf t}) = r^m = \delta_{m1}
\ee
and particular $\Psi^m_1$ (in principle (\ref{recur})
together with (\ref{n0}) say only that $\Psi^m_1$ are linear
functions of times -- with arbitrary coefficients)
\be
\label{n1}
\Psi^m_1({\bf t}) = \sum_l\eta_{kl}t_l
\ee
where $\eta_{kl}=(\f_1)_{kl}$ is ``topological metric'' supposed to be
constant, say,
in the Landau-Ginzburg case of sect.~\ref{ss:lg}, there is a simple
relation between prepotential $\F$
and first coefficients $\Psi^m_0$, $\Psi^m_1$, $\Psi^m_2$ and $\Psi^m_3$ of
this particular solution.

Indeed, the $n=2$ equations after substituting (\ref{n1})
gives
\be
\label{psi2}
\d_i\d_j\Psi^m_2 = \sum_k C_{ij}^k\eta_{km}
\ee
which is solved by
\be
\label{n2}
\Psi^m_2 = {\d \F\over\d t_m}
\ee
where $\F $ is prepotential. Going further, one gets
\be
\label{toprr}
\d_i\d_j\Psi^m_3 = \sum_k C_{ij}^k{\d^2\F\over\d t_m\d t_k}
\ee
which is nothing but the so called ``topological recursion relation''
\be
\langle\sigma_1(\phi_m)\phi_i\phi_j\rangle = \sum_k C_{ij}^k
\langle\phi_m\phi_k\rangle
\ee
for
the ``first descendants'', so one may identify
\be
\Psi^m_3 \equiv \langle\sigma_1(\phi_m)\rangle = {\d\F\over\d T_{1,m}}
\ee
Now, one may easily notice that prepotential $\F$ can be expressed as
\be
\label{AKV}
\F = {1\over 2}\sum_{i,j}\eta_{ij}\left(\Psi_1^i\Psi_2^j -
\Psi_0^i\Psi_3^j\right)
\ee
Upon (\ref{n0}), (\ref{n1}) and (\ref{n2}) this turn into "renormalization
group" like equation
\be
\label{rglike}
\F = {1\over 2}\left(\sum_k  t_k{\d\F\over\d t_k} -
\Psi^1_3\right)
\ee
or
\be
\label{rg}
2\F - \sum_k t_k{\d\F\over\d t_k} = - \Psi^1_3 = - {\d\F\over\d T_{1,1}}
\ee
(the last equality has just a symbolic meaning).
From (\ref{rg}) it follows further that
\be
{\d\F\over\d t_i} = \sum_kt_k{\d^2\F\over\d t_i\d t_k} -
{\d\Psi^1_3\over\d t_i}
\\
{\d^2\Psi^1_3\over\d t_i\d t_j} = \sum_k
t_k{\d^3\F\over\d t_i\d t_j\d t_k}
\ee
The last equality can be rewritten using (\ref{n2}) and (\ref{psi2}) as
\be
{\d^2\Psi^1_3\over\d t_i\d t_j} = \sum_{k,n}C_{ij}^k \eta_{kn}t_n
= \sum_k C_{ij}^k {\d^2\F\over\d t_k\d t_1}
\ee
which turns to be a particular case of "topological recursion relations"
(\ref{toprr}). Thus, the proof of (\ref{AKV}) is rather trivial and does
{\em not} require anything except for substitution of the
expansion (\ref{expxi}) and (\ref{expxi1}) into (\ref{aulixi}).

\subsection{Linear problem and Landau-Ginzburg versus
Seiberg-Witten cases}

Formula (\ref{aulixi}) is in fact equivalent to special dependence of
the flat section $\Psi (t_1,t_2,\dots)\sim
e^{\zeta t_1}\Psi(\zeta;t_2,\dots)$ upon the first time, motivated, for
example, by Kontsevich model. There is no need to require this in general,
so one can consider the auxiliary linear problem in more general form
\be
\label{auli}
\left({\d^2\over \d t_i\d t_j} - \sum_k C_{ij}^k
{\d^2\over \d t_k\d t_1}\right)\Psi = 0
\ee
Consider the Landau-Ginzburg case where the
coefficients $C_{ij}^k$ are provided by (\ref{ringw}).
Another simple exercise is to check that (\ref{auli}) processes
a solution
\be
\label{psigamma}
\Psi ({\bf t};\Gamma) = \oint_\Gamma f(W(\lambda;{\bf t}))d\lambda
\ee
for {\em any} function $f(W)$ and any non-contractible contour $\Gamma$.
Indeed,
\be
\label{calc}
\left({\d^2\over \d t_i\d t_j} - \sum_k C_{ij}^k
{\d^2\over \d t_k\d t_1}\right)f(W) = \\ =
f'(W)\left({\d^2 W\over \d t_i\d t_j} - \sum_k C_{ij}^k
{\d^2 W\over \d t_k\d t_1}\right) +
f''(W)\left({\d W\over \d t_i}{\d W\over\d t_j} - \sum_k C_{ij}^k
{\d W\over \d t_k}{\d W\over\d t_1}\right)
\ee
Using (\ref{d1}) and (\ref{ringw}) the second term in the r.h.s. of
(\ref{calc}) gives $f''(W)R_{ij}W'$, while the first is just
$f'(W){\d^2 W\over \d t_i\d t_j}$, again due to (\ref{d1}). Combining
this together and using (\ref{gm}) one finally gets
\be
f'(W){\d^2 W\over \d t_i\d t_j} + f''(W)R_{ij}W' =
f'(W)\d_\lambda R_{ij} + R_{ij}\d_\lambda f'(W) =
{d\over d\lambda}\left(R_{ij}f'(W)\right)
\ee
what proves that integral (\ref{psigamma}) solves (\ref{auli}). It is
important that it works for {\em any} function $f(W)$.

Are there any other restrictions to the function $f(W)$? One may only add
that, to relate this with the WDVV equations
there should be exactly $2(N-1)$ independent integrals
(\ref{psigamma}) and, that the derivatives
\be
\label{hol1}
{\d f(W)\over \d t_i} = f'(W){\d W\over \d t_i} = f'(W)\d_\lambda W^{i/N}_+
\ee
should be holomorphic, since only then it is possible to come to {\em new}
set of independent variables $a_i = \Psi({\bf t},\Gamma_i)$, $i=1,\dots,N-1$
and
to prove that there exists a function ${\cal F}({\bf a})$ such that
\be
a^D_i = \Psi({\bf t},\Gamma_{2N-1-i}) = {\d\F\over\d a_i}
\ee
But this is almost reformulation of the case of Seiberg and Witten (cf. with
(\ref{aad})).
Equivalently one may say that if there are $2(N-1)$ independent solutions
(\ref{psigamma}) then the equations (\ref{auli}) may be rewritten as
\be
\sum_{n,m}{\d^2 a^D\over\d a_n\d a_m}{\d a_n\over\d t_i}{\d a_m\over\d t_j} =
\sum_{k,l,s}C_{ij}^k {\d^2 a^D\over\d a_l\d a_s}
{\d a_l\over\d t_k}{\d a_s\over\d t_1}
\ee
with $a^D = \Psi({\bf t},\Gamma)$ corresponding to the ``rest of'' the contours
$\Gamma_s$ with $s=N,\dots,2N-2$ if $a_n = \Psi({\bf t},\Gamma_n)$ with
$n=1,\dots,N-1$. In different words
\be
{\d^2 a^D\over\d a_n\d a_m} = \sum C_{nm}^l({\bf a})\eta(a^D,l)
\ee
with
\be
C_{nm}^l({\bf a}) = \sum_{i,j,k}{\d t_i\over\d a_n}{\d t_j\over\d a_m}
C_{ij}^k({\bf t}){\d a_l\over\d t_k}
\\
\eta(a^D,l) = \sum_s {\d a_s\over\d t_1}{\d^2 a^D\over\d a_l\d a_s}
\ee
There is absolutely no mystery in the above formulas for transformation of
the second derivatives, one should just stay that one should take
{\em exactly} $(N-1)$ solutions (\ref{psigamma}) as new independent
coordinates. For particular choice of function $f(W)$ one may get from these
formulas relation between the transformations (\ref{ceta}) for the
Landau-Ginzburg and
Seiberg-Witten cases (see, for example, \cite{ItoYang,luukmart}).

As for solutions to (\ref{hol1}), say, any ``hyperelliptic integral''
\be
f(W) = \int^W {dX\over\sqrt{{\rm Pol} (X)}}
\ee
provides such a solution. There should be also solutions of this kind
related to non-hyperelliptic curves and they could lead to the new solutions
of the WDVV equations (\ref{WDVV}), we are going to discuss this problem
in a separate publication \cite{HMM}.

\subsection{Auxiliary linear problem and Hirota equations}

In order to understand better the relation between auxiliary linear problem
and formulas like (\ref{ceta}) let us notice, that in the case of
infinitely many variables
the auxiliary linear problem (\ref{auli}) can be rewritten with the help of
generating operators (\ref{eq:4}) - (\ref{eq:14}) as
\be
\label{auliinf}
\left( D_1D_2 - \sum_k C^k(z_1,z_2){\d\over\d t_k}
{\d\over\d t_1}\right)\Psi = 0
\ee
with the structure constants $C^k(z_1,z_2)$ given by (\ref{eq:142}).
The aim of this section is to compare eq.~(\ref{auliinf}) with the Hirota
equation (\ref{eq:7}).

Let us suppose that tau-function $\F$ depends also upon some additional
parameters $T$, {\it i.e.} $\F = \F({\bf t};T)$ whose meaning is arbitrary in this
context, though it is natural to suppose that these parameters would
correspond to the descendants we already discussed in sect.~\ref{ss:AKV}.
We also imply that our usual tau-function or prepotential $\F = \F(t;T_0)$
is considered as some fixed values of the extra parameters $T=T_0$ and we
may also consider a perturbation
\be
\label{pertFT}
\F(t;T) = \F(t,T_0+\epsilon) = \F(t;T_0) + \epsilon{\d\F\over\d T} + \dots
\equiv \F(t;T_0) + \epsilon\Psi (t;T_0) + \dots
\ee
Now, it is easy to demonstrate that $\Psi$ is a solution to auxiliary linear
problem (\ref{auliinf}). Indeed, consider the dispersionless Hirota
relations in the form (\ref{eq:7}) and substitute there (\ref{pertFT}). One
gets
\be
\label{eauli}
(z_1-z_2)\left(1 - e^{D_1D_2\F}e^{\epsilon D_1D_2\Psi + \dots}\right) =
(D_1 - D_2)\d_{t_1}\F + \epsilon (D_1 - D_2)\d_{t_1}\Psi + \dots
\ee
The zeroth order in $\epsilon $ terms in (\ref{eauli}) give rise to
"old" Hirota relations
(\ref{eq:7}) for the nonperturbed function $\F(t;T_0)$. The linear in
$\epsilon$ terms combine into
\be
D_1D_2\Psi = e^{-D_1D_2\F}{D_1-D_2\over z_1-z_2}\d_{t_1}\Psi
\ee
or exactly (\ref{auliinf}) provided by (\ref{eq:142}).

\section{Conclusion}

The aim of this paper was to consider some interesting properties of the
solutions to associativity equations (\ref{WDVV}) which arise often in the
context of effective actions for various physical theories. We have tried to
demonstrate that these properties are very universal and that reflects
probably the universality of the equations (\ref{WDVV}) themselves.

Equations (\ref{WDVV}) are consequence of an associative algebra, which, as
we tried to show exists very often and the relation between its structure
constants and third derivatives of some function $\F$ (\ref{ceta}). The last
requirement is much less trivial and strongly depends, in contrast to
associative algebra itself, upon the choice of the basis; in many
interesting physical examples this relation follows from localization in the
form of residue formula (\ref{res}). We have concentrated on this important
class of solutions whose universality maybe very close, if not identical, to
the universality of WDVV equations themselves.

Let us now list the main messages made above in the text:

\begin{itemize}

\item Having residue formula the rest of the proof of WDVV equations is
rather simple and based only on degeneracy and matching condition. Two
different new families of the solutions to WDVV equations, compare to
two-dimensional topological theories, exactly give two possible ways of
at least naive violating of matching condition.

\item The proof does not require {\em any} additional requirements or
structures. From this point of view all extra requirements onto solutions to
WDVV equations coming from topological theories are inessential. Nothing
depends on the constancy of topological "metric" and the notion of "metric"
itself is absolutely unclear in more general set-up for associativity
equations.

\item Moreover, when it is clear that underlying WDVV equations geometry
can be identified with special K\"ahler geometry, the covariance of
equations (\ref{WDVV}) requires real metric {\em not} to appear in
associativity equations. WDVV equations are covariant, but this is not true
for the class of their "topological " solutions.

\item One may hope that real geometric sense of the associativity equations
may be related to the fact that they come from certain algebraic relations
for the second derivatives of the function $\F$. A particular example comes
from dispersionless quasiclassical hierarchies, whenever a function $\F$ is
a solution to algebraic dispersionless Hirota relations it solves the WDVV
equations (\ref{WDVV}).

\end{itemize}

\section*{Acknowledgements}

I am grateful to A.Boyarsky, H.Braden, B.de~Wit, L.Hoevenaars, I.Krichever,
A.Losev, A.Mironov, A.Morozov, S.Natanzon, V.Rubtsov, O.Ruchayskiy,
P.Wiegmann and A.Zabrodin
for collaboration and many very fruitful discussions.
The work was
partially supported by RFBR grant
No.~01-01-00539, INTAS grant No.~99-01782, CRDF grant No.~RP1-2102
(6531) and the grant for the support of scientific schools
No.~00-15-96566.
I am also grateful to the organizers of the workshops
in SISSA, Chennai and Allahabad for nice atmosphere and hospitality.
%

\end{document}